\documentclass[useAMS,usenatbib]{mn2e}
\usepackage{graphicx}
\usepackage{amssymb}
\usepackage{txfonts}
\usepackage{epstopdf}
\usepackage{amsfonts}
\usepackage{natbib}
\usepackage{epsfig}
\usepackage{multicol}
\usepackage{epsfig}

\usepackage{mncite}

\def\tc{t_{\mathrm{cool}}}
\def\tth{t_{\mathrm{therm}}}
\def\dSoS{\delta \Sigma/\bar{\Sigma}}

\begin{document}


\title{Characterising the Gravitational Instability in Cooling Accretion Discs}

\author[P.~Cossins, G.~Lodato \& Cathie Clarke] 
{
\parbox{5in}{Peter Cossins$^1$, Giuseppe Lodato$^1$ and C. J. Clarke$^2$}
\vspace{0.1in} 
 \\ $^1$ Department of Physics \& Astronomy, University of Leicester,
 Leicester LE1 7RK, UK\\
 $^2$ Institute of Astronomy, Madingley Road, Cambridge, CB3 0HA
}

\maketitle


\begin{abstract}
  In this paper we perform a systematic analysis of the structure induced by
  the onset of gravitational instabilities in cooling gaseous accretion discs.
  It is well known that for low enough cooling rates the disc reaches a
  quasi-steady configuration, with the instability saturating at a finite
  amplitude such that the disc is kept close to marginal stability. We analyse
  the dependence of the saturation amplitude on the imposed cooling rate, and
  we find that it scales with the inverse square root of the cooling parameter
  $\beta = t_{\mathrm{cool}} / t_{\mathrm{dyn}}$.  This indicates that the
  heating rate induced by the instability is proportional to the energy
  density of the density waves excited by the disc self-gravity. In
  particular, we find that at saturation the energy dissipated per dynamical
  time by weak shocks due to the gravitational perturbations is of the order
  of 20 per cent of the wave energy. We further perform a Fourier analysis of
  the disc structure, and subsequently determine the dominant radial and
  azimuthal wavenumbers of the density waves. While the number of spiral arms
  (corresponding to the azimuthal wavenumber) is fairly constant with radius,
  we find that the disc displays a range of radial wavenumbers whose mean
  increases with increasing radius.  The dominant modes closely match the
  locally most unstable wavelength as predicted by linear perturbation
  analysis. As a consequence, we demonstrate numerically that the density
  waves excited in relatively low mass discs $M_{\mathrm{disc}} / M_{*} \sim
  0.1$ are always close to co-rotation, deviating from it by approximately 10
  per cent.  This result can be understood in terms of the constancy of the
  Doppler-shifted phase Mach number of the flow -- the pattern speed
  self-adjusts so that the flow into spiral arms is always sonic. This has
  profound effects on the degree to which the extraction of energy and angular
  momentum from the mean flow through density waves can be modelled as a
  viscous process.  Our results thus provide (a) a detailed description of how
  the self-regulation mechanism is established for low cooling rates, (b) a
  clarification of the conditions required for describing the transport
  induced by self-gravity through an effective viscosity, (c) an estimate of
  the maximum amplitude of the density perturbation before fragmentation takes
  place, and finally (d) a simple recipe to estimate the density perturbation
  in different thermal regimes.
\end{abstract}


\footnotetext{E-mail: peter.cossins@astro.le.ac.uk}


\section{Introduction}
\label{intro}

In cold, relatively massive accretion discs the effects of the disc
self-gravity can become dynamically important, and gravitational instabilities
may play a major role in determining the long-term evolution of the disc.  On
the one hand the instability can induce fragmentation of the disc,
potentially resulting in the formation of massive stars in AGN accretion discs
\citep{Nayakshin07} or, in the case of proto-planetary discs, low-mass
companions \citep{StamatellosHW07} or even giant planets \citep{Boss97,Boss98}.
On the other hand, the instability is very effective in transporting and
redistributing angular momentum within the disc \citep{LodatoR04,LodatoR05}
and might therefore promote accretion in a variety of different contexts and
scales, from supermassive black hole growth \citep{ShlosmanBF90,LodatoN06} to
star formation \citep{VorobyovBasu05}.    

The balance between the heating provided through such gravitational
instabilities and the cooling as energy is radiated away is known 
to have major implications for the fate of the disc.  Fragmentation into bound
objects occurs when the cooling dominates, whereas the disc can persist in a
quasi-equilibrium marginally stable condition if the cooling rate is
sufficiently low \citep{Gammie01, RiceLA05, LodatoR04}.  

Such a quasi-steady state is characterised by the presence of spiral density
waves, which transport energy and angular momentum through the disc.  Where
the discs are weakly ionised, and therefore where the magneto-rotational
instability is likely to be ineffective, this  angular momentum transport
mechanism may be the primary driver of accretion.   Additionally, these waves
extract rotational energy from the disc, some of which is then returned as
heat as the waves steepen into shocks. This heat then stabilises the disc
against further gravitational collapse into bound fragments.  As the amplitude
of the waves increases, so too does the wave energy density, increasing the
reservoir of energy available to be returned to the disc as heat.  However,
numerical experiments \citep{Gammie01, RiceLA05} have shown that
if the cooling time $\tc$ is less than a few times the local dynamical time
$\Omega^{-1}$, this feedback process breaks down and fragmentation ensues.  The
quasi-steady marginal stability state therefore represents a restricted regime
of dynamic thermal equilibrium, where the heating through gravitational
instabilities is balanced by the cooling rate. 

In this paper we seek to characterise the relationship between the strength of
the cooling and the amplitude of the spiral density waves excited within the
disc through self-gravity, while remaining within this dynamic thermal
equilibrium state. To this end we use a Smoothed Particle Hydrodynamics (SPH)
code to run global numerical simulations of self-gravitating gaseous discs.
From such controlled numerical experiments, we measure the amplitude of the
density perturbations over a range of cooling times, down to the limit where
the disc fragments, and to investigate the overall structure formation.
Fourier analysis allows us to characterise the mode spectra and pattern speeds
associated with this structure, and to associate the dynamics of the spiral
density waves excited through self-gravity with the thermodynamics of the disc
self-regulation process.  We begin by discussing some relevant properties of
density waves and transport processes in gaseous discs in Section 2, and link
this to the disc thermodynamics in Section 3.  Details of the numerical
simulations and initial conditions are given in Section 4, and the results of
these simulations are presented in Section 5.  In Section 6 we discuss the
implications of these results and the conclusions that may be drawn from them.


\section{Dynamics of Self-Gravitating Gaseous Discs}
\label{dynamics}

In the following two sections we derive analytically the basic relations
that link the disc quantities in the presence of perturbations to the induced
transport properties of self-gravitating discs. These standard results have
been mostly derived in the context of stellar dynamics \citep{Shu70,Bertin00,
  BinneyTremaine2e}.  In the context of gaseous (accretion) discs,
qualitatively similar (but rather more involved) analyses have been presented
in \citet{BalbusPap99} and \citet{Balbus03}.   

\subsection{Dispersion Relations for Self-Gravitating Discs}
The linear development of the gravitational instability is usually described
by the dispersion relation $D(\omega,k,m)$ between the wave (angular)
frequency $\omega$ and the radial and azimuthal wavenumbers of excited modes,
$k$ and $m$ respectively.  For an infinitesimally thin (i.e. 2-dimensional)
disc, this can be obtained using the standard WKB method of perturbation
analysis, and in the tight-winding limit (where the radial wavelength is
small in comparison to the azimuthal wavelength, $|kR| \gg m$, with $R$ the
cylindrical radius) we have   
\begin{equation}
  D(\omega,k,m) = (\omega - m\Omega)^{2} - c_{\mathrm{s}}^{2}k^{2} + 2 \pi G
  \Sigma |k| - \kappa^{2} = 0,
  \label{dispersionrelationD}
\end{equation}
where $\Omega(R)$ is the angular speed, and $\kappa(R)$ is the epicyclic
frequency, $c_{\mathrm{s}}(R)$ is the sound speed and $\Sigma(R)$ is the
surface density of the umperturbed disc \citep{BinneyTremaine2e}.  Introducing
the pattern speed of the wave $\Omega_{\mathrm{p}}(R)$, where $\omega =
m\Omega_{\mathrm{p}}$, we may express this as follows;
\begin{equation}      
  m^{2}(\Omega_{\mathrm{p}} - \Omega)^{2} = c_{\mathrm{s}}^{2}k^{2} - 2 \pi G
  \Sigma |k| + \kappa^{2}.
\label{dispersionrelation}
\end{equation}

The well known stability criterion for axisymmetric disturbances
\citep{Toomre64} 
\begin{equation} 
  Q = \frac{c_{\mathrm{s}}\kappa}{\pi G \Sigma}>1
  \label{Q} 
\end{equation}
identifies the region of parameter space where the RHS of
Eq.(\ref{dispersionrelation}) is positive definite, and therefore where the
disc is stable at all wavelengths.  For the case of an unstable disc, the most
unstable wavelength $k_{\mathrm{uns}}$ (i.e. where the RHS of
Eq.(\ref{dispersionrelation}) is at a minimum) occurs at a radial wave-number
\begin{equation}
  k_{\mathrm{uns}} = \frac{\pi G \Sigma}{c_{\mathrm{s}}^{2}} .
  \label{kuns}
\end{equation}
For a marginally stable disc where $Q \approx 1$ we would therefore expect
only modes with $k \approx k_{\mathrm{uns}}$ to be excited, and in this
instance, Eq.(\ref{dispersionrelation}) tells that $\Omega_{\mathrm{p}} \approx
\Omega$, i.e. all excited modes are expected to be close to co-rotation. Note
that the most unstable wave-number is exactly the inverse of the disc
thickness for a self-gravitating disc 
\begin{equation}
  H = \frac{c_{\mathrm{s}}^{2}}{\pi G\Sigma} \approx Q
  \frac{c_{\mathrm{s}}}{\Omega}, 
  \label{H}
\end{equation}
where the last approximation is exact for a Keplerian rotation curve. 

\subsection{The Stress Tensor}
In non-axisymmetric self-gravitating discs, torques arising due to the
perturbed gravitational potential play an important role in transporting both
energy and angular momentum.  

Traditionally, viscous accretion discs are described by the $\alpha$-formalism
of \citet{ShakSun73}, where (for an infinitesimally thin disc) the only
non-vanishing component of the vertically integrated stress tensor
$\mathbf{T}$  is the shear term, defined as 
\begin{equation}
  T_{R\phi} = \alpha \Sigma c_{\mathrm{s}}^{2}  \frac{\mathrm{d} \ln
    \Omega}{\mathrm{d} \ln R},
  \label{T}
\end{equation}
where $\alpha \lesssim 1$ is a dimensionless parameter which measures the
viscosity.  In this case the disc stress is linked to the local pressure
$\Sigma c_{\mathrm{s}}^{2}$, and indicates that the $\alpha$-formalism is
fundamentally a local relationship.  Furthermore, since $\alpha$ is not
necessarily constant, this represents a completely general description for any
purely local process.  Also note that for Keplerian rotation, $\mathrm{d} \ln
\Omega / \mathrm{d} \ln R = -3/2$, implying that the stress is negative,
i.e. it acts to oppose rotation, and therefore allows for inward accretion
flows. 

Accretion disc theory identifies the origin of the stress with 
torques arising due to perturbations in a ``turbulent'' disc.  Crucially for
the gas dynamics, these perturbations manifest themselves as fluctuations in
the mean flow velocity, the gravitational potential and the magnetic field
threading the disc.  For non-magnetised self-gravitating discs such as we
consider here, the stress tensor can therefore be broken down into a Reynolds
stress term, associated with velocity fluctuations, and a gravitational stress
term, associated with fluctuations in the gravitational potential.  The
Reynolds stress term $T_{R\phi}^{\mathrm{Reyn}}$ is such that 
\begin{equation}
  T_{R\phi}^{\mathrm{Reyn}} = \Sigma \langle\delta v_{R} \delta v_{\phi}\rangle,
  \label{TReyn}
\end{equation}
(where $\delta v_{R}, \delta v_{\phi}$ are the velocity fluctuations about the
mean flow velocity in the $R$ and $\phi$ directions respectively and the
brackets indicate azimuthal averaging) and the gravitational stress term
$T_{R\phi}^{\mathrm{grav}}$ is given by \citet{LyndenBellKalnajs72} as
\begin{equation}
  T_{R\phi}^{\mathrm{grav}} = \int \left\langle\frac{g_{R}g_{\phi}}{4 \pi
      G}\right \rangle dz
  \label{Tgrav}
\end{equation}
where again the $g_{R},g_{\phi}$ are the accelerations due to the perturbed
gravitational potential of the disc in the $R$ and $\phi$ directions
respectively.

The viscous torque per unit area $\mathbf{\dot{L}}_{\alpha}$ is related to
the vertically integrated stress tensor $\mathbf{T}$ through
\begin{equation}
  \mathbf{\dot{L}}_{\alpha} = R \nabla \cdot \mathbf{T},
\end{equation}
which in turn yields
\begin{equation}
  \dot{\mathcal{L}}_{\alpha} = \frac{\partial}{\partial R} (R T_{R\Phi}) 
  \label{viscoustorque}
\end{equation}
as the only non-zero component of the torque.  The power per unit surface
$\dot{\mathcal{E}}_{\alpha}$ produced by this viscous torque is then given
simply by \citep{FrankKR,Pringle81}
\begin{equation}
  \dot{\mathcal{E}}_{\alpha} = \Omega \dot{\mathcal{L}}_{\alpha},
  \label{localpower}
\end{equation}
where the subscript $\alpha$ indicates that this relation is expected for a
viscous disc.  Equation (\ref{localpower}) therefore links the transport of
angular momentum and the associated rate of work done by torques in the case
of a local process, as historically modelled via the $\alpha$-viscosity
parameter.  

\subsection{Wave Energy and Angular Momentum Densities}
In this section we turn our attention to the transport of energy and
angular momentum through the propagation of the spiral density waves
expected to arise in a self-gravitating disc.  To this end it is convenient to
introduce the wave action density $\mathcal{A}$ for density waves.  Within the
WKB approximation used to derive the dispersion relation given in
Eq.(\ref{dispersionrelation}),  the wave action per unit surface is given by
\citep{Toomre69,Shu70,FanLou99}  
\begin{equation}
  \mathcal{A} = \frac{m(\Omega_{\mathrm{p}} - \Omega)}{8\pi^{2} G^{2} \Sigma}
  |\delta \Phi|^{2},
  \label{waveaction}
\end{equation}
where $\delta\Phi$ is the perturbed gravitational potential, itself given
by 
\begin{equation} 
  \delta \Phi = -\frac{2\pi G \delta \Sigma}{|k|},
  \label{perturbedphi}
\end{equation}
where $\delta \Sigma$ is the surface density perturbation.  
The wave energy per unit surface $\mathcal{E}_{\mathrm{w}}$ and the 
wave angular momentum per unit surface $\mathcal{L}_{\mathrm{w}}$ 
are obtained in a straightforward way from the wave action through the 
standard wave dynamics relations \citep{Bertin00,Shu70},
\footnote{Note that this system of equations is analogous to
  that found in quantum mechanics for an harmonic oscillator; from the
  quantum of action $\hbar$, the quantized energy $E$ and angular momentum
  $S$ are found via $E = \hbar \omega$ and $S = \hbar m$, where $\omega$ and
  $m$ are the angular frequency and spin quantum number respectively.} 

\begin{equation}
    \mathcal{E}_{\mathrm{w}} = \omega \mathcal{A} = m \Omega_{\mathrm{p}}
    \mathcal{A}, 
     \label{wavedensities}
\end{equation}
\begin{equation}
    \mathcal{L}_{\mathrm{w}} = m \mathcal{A}.
    \label{wavedensities2}
\end{equation}

Combining the first of these relations with Eqs.(\ref{waveaction}) and
(\ref{perturbedphi}) we obtain
\begin{equation}
  \mathcal{E}_{\mathrm{w}} = \frac{\Sigma v_{\rm p} \tilde{v}_{\rm p}}{2}
  \left(\frac{\delta \Sigma}{\Sigma} \right)^{2},
  \label{waveenergydensity}
\end{equation}
where 
\begin{equation}
  v_{\mathrm{p}} = m \Omega_{\mathrm{p}} / k
\label{phasespeed}
\end{equation}
\begin{equation}
 \tilde{v}_{\mathrm{p}} = m(\Omega_{\mathrm{p}} - \Omega) / k
\label{Dopplerphasespeed}
\end{equation}
are the radial phase speed and Doppler-shifted radial phase speed of the wave
respectively.  Note that Eqs.(\ref{waveenergydensity}) and
(\ref{Dopplerphasespeed}) together explain why self-induced density waves are
launched at co-rotation -- since the energy density changes sign at
co-rotation, waves that propagate away from co-rotation extract no net energy
from the flow.

Looking again at Eqs.(\ref{wavedensities}) and (\ref{wavedensities2}), we see
that the  relationship between the energy per unit surface and the angular
momentum per unit surface in a density wave is given by
\begin{equation}
  \mathcal{E}_{\mathrm{w}} = \Omega_{\mathrm{p}} \mathcal{L}_{\mathrm{w}}.
  \label{wavetrans}
\end{equation}

In the case of quasi-stationary waves (where $\Omega_{\mathrm{p}}$ is constant)
propagating in a disc in dynamic thermal equilibrium, the rate at which energy
is lost per unit surface due to cooling must be balanced by the
power per unit surface $\dot{\mathcal{E}}_{\mathrm{w}}$ dissipated by the
waves. In order to maintain the amplitude of the wave, the instability has to
keep extracting energy and angular momentum from the background flow.  The
fluxes of energy (angular momentum) carried by the wave are simply
$\mathcal{E_{\mathrm{w}}}$ ($\mathcal{L_{\mathrm{w}}}$) times the local
group velocity.  Hence when a wave dissipates it adds energy and angular
momentum to the flow in the ratio of $\mathcal{E_{\mathrm{w}}}$ to
$\mathcal{L_{\mathrm{w}}}$, i.e. in the ratio $\Omega_{\mathrm{p}}$.  We
therefore conclude that
\begin{equation} 
  \dot{\mathcal{E}}_{\mathrm{w}} = \Omega_{\mathrm{p}}
  \dot{\mathcal{L}}_{\mathrm{w}}. 
  \label{wavepower}
\end{equation}

Equation (\ref{wavepower}) is the wave analogue of Eq.(\ref{localpower}).
Comparing these two equations, we note a fundamental difference with respect
to the viscous model -- for a given torque $\dot{\mathcal{L}}$, waves extract
energy from the flow at a rate proportional to the wave pattern speed
$\Omega_{\mathrm{p}}$, whereas the rotation speed $\Omega$ is the underlying
rate in the local (viscous) case.

\citet{BalbusPap99} have noted similarly that in general, energy
transport through the gravitational instability cannot be described purely in
viscous terms, and indeed this is only possible at co-rotation, when
$\Omega_{\mathrm{p}} = \Omega$.  This can be readily understood if we consider
that the wave energy per unit surface $\mathcal{E}_{\mathrm{w}}$
(Eq.(\ref{waveenergydensity})) can be decomposed into
two separate terms, as follows;
\begin{equation}
  \begin{array}{lcl}
    \displaystyle
    \mathcal{E}_{\mathrm{w}} & = &\displaystyle
    \frac{\Sigma}{2}\frac{m^{2}}{k^{2}} (\Omega_{\mathrm{p}} - \Omega)^{2}
    \left(\frac{\delta \Sigma}{\Sigma} \right)^{2} \\     
    &  +  & \displaystyle \frac{\Sigma}{2}\frac{m^{2}}{k^{2}}     \Omega
    (\Omega_{\mathrm{p}} - \Omega) \left(\frac{\delta \Sigma}{\Sigma}
    \right)^{2}. \\       
    \label{energysplit}
  \end{array}
\end{equation}
The second term, equal to the angular momentum per unit surface times the
rotation speed $\Omega$, is a local energy transport term (cf.
Eq.(\ref{localpower})) and can therefore be represented using the
$\alpha$-formalism.  The first term however, equal to the same angular
momentum term times $\Omega_{\mathrm{p}} - \Omega$, is a non local term. In
fact the energy flux associated with this non-local transport term is is
precisely that identified by \citet{BalbusPap99} as an ``anomalous flux'',
preventing self-gravitating discs from acting as pure $\alpha$-discs.

We see that far from co-rotation, where $\Omega \neq \Omega_{\mathrm{p}}$,
this term becomes significant, and thus non-local (global) transport becomes
important within the disc.  For trailing waves launched at co-rotation, the
direction of energy (and angular momentum) transport is outwards throughout
the disc (i.e. inward travelling waves with negative energy density inside
co-rotation and outward travelling waves with positive energy density outside
co-rotation), and thus the dissipation of such waves effects a net outward
transport of energy (and angular momentum).  If such a wave dissipates at
large radius (where $\Omega_{\mathrm{p}} \gg \Omega$) then the ratio in which
energy and angular momentum are added to the disc (Eq.(\ref{wavepower})) is
significantly greater than the equivalent ratio in the viscous case
(Eq.(\ref{localpower})).  Consequently, under such conditions the energy
dissipated at large radii in a steady state disc with wave transport can
significantly exceed that dissipated in an equivalent viscous disc -- with
this extra energy being extracted by the wave from deep in the potential and
transported to large radii \citep{LodatoBertin01,BertinLodato01}.  However, if
waves instead dissipate close to co-rotation, the wave transport is dominated
by the local term in Eq.(\ref{energysplit}); since energy and angular
momentum transport are exchanged with the disc in roughly the same ratio as
for a viscous process, then in this regime the $\alpha$-formalism is a good
approximation to the actual transport properties of the disc.

From Eq.(\ref{energysplit}) it is therefore possible to quantify a non-local
transport fraction $\xi$, from the ratio of the two terms on
the RHS, such that    
\begin{equation}   
  \xi = \left| \frac{\Omega - \Omega_{\mathrm{p}}}{\Omega} \right|.
  \label{globaltransportfraction}
\end{equation}
Thus in order to assess the importance of non-local effects, we need to know
the relationship between the angular frequency $\Omega$ and the pattern speed
$\Omega_{\rm p}$.  In section \ref{locality} we use the dispersion relation
along with information extracted through Fourier analysis in order to estimate
the pattern speed, and hence to evaluate $\xi$ directly. 


\section{Simulating the Disc Thermodynamics}
\label{thermo}

Realistically simulating the thermodynamics of accretion discs is a complex
undertaking and as such has received much attention, from the opacity-based
treatment employed by \citet{JandG03} through to the various convective
and radiative transfer models of \citet{Boss04}, \citet{Boleyetal07},
\citet{Mayeretal07}, \citet{StamatellosW07} and \citet{StamatellosW08}, the
latter two of which also account for heating from the central star.

However, in this paper, one of our aims is to investigate the relationship
between the properties of the density perturbations and the rate at which the
disc cools. This purpose is served most readily by \textit{imposing} a known
cooling rate against which the heating rate and subsequent disc structure may
be easily correlated. It is therefore not necessary to consider the exact
physics of the cooling regimes found in astrophysical discs, and hence we can
use a cooling law for the heat loss rate per unit mass $Q^{-}$, such
that    
\begin{equation}
  Q^{-} = -\frac{u}{\tc},
  \label{Qminus}
\end{equation}
where $u$ is the internal energy per unit mass and where the details of the
cooling function (and possibly of additional \textit{external} heating) are
absorbed into the simple parameter $t_{\mathrm{cool}}$. As long as such a
characteristic timescale can be defined, it is therefore possible to use this
formalism to represent a wide range of cooling mechanisms.  Within this paper,
we use a fixed ratio between the local dynamical and cooling timescales, such
that $\beta = \Omega \tc$ is constant. This form of cooling has been used
extensively in simulations of discs in various contexts, for example
\citet{Gammie01}, \citet{LodatoR05}, \citet{HobbsN08}, and has proven useful
in elucidating the properties of the gravitational instability in controlled
numerical experiments.      

In terms of the heating from the gravitational instability, we noted in
the previous section that density waves extract energy from
the disc.  In addition to compression heating, in the case where the pattern
speed differs from the rotation speed by more than the local sound speed,
these waves will steepen into shocks, liberating further heat into the disc.
We expect the rate at which energy is added to the disc to scale with
the energy of the wave and the local dynamical timescale, and we can then
express the heating rate per unit mass due to the instability as
\begin{equation}
  Q^{+}= \frac{1}{\Sigma}\epsilon \Omega |\mathcal{E}_{\mathrm{w}}|
  = \epsilon \Omega c_{\mathrm{s}}^{2}  \frac{\mathcal{M}
    \widetilde{\mathcal{M}}}{2} \left(\frac{\delta \Sigma}{\Sigma} \right)^{2},
  \label{Qplus}
\end{equation}
where we define the radial phase and Doppler shifted phase Mach numbers to be
$\mathcal{M} = |v_{\mathrm{p}}| / c_{\mathrm{s}}$ and $\widetilde{\mathcal{M}}
= |\tilde{v}_{\mathrm{p}}| / c_{\mathrm{s}}$ respectively. In
Eq. (\ref{Qplus}) we have also introduced a dimensionless proportionality
factor $\epsilon$, hereinafter referred to as the heating factor. If the
relationship between the pattern speed and the angular speed is self-similar
(i.e., it does not vary across the disc), we expect $\epsilon$ to be simply a
constant, independent of radius. 

Once the gravitational instability has been instigated and has subsequently
saturated, we may consider the disc to be in dynamic thermal equilibrium, such
that the energy released through wave-driven shock and compression heating is
balanced by the imposed cooling, i.e. $Q^{-} + Q^{+} = 0$. Recalling that $u =
c_{\mathrm{s}}^{2} / \gamma(\gamma - 1)$, we can equate equations (\ref{Qplus})
and (\ref{Qminus}) and thereby determine the following relationship between
the amplitude of the density perturbations and strength of the cooling, as
measured by the $\beta$ parameter;    
\begin{equation}
  \left( \frac{\delta \Sigma}{\Sigma} \right)^{2} =  \frac{2}{\epsilon
    \beta}\frac{1}{   \gamma (\gamma - 1)} \left(\frac{1}{\mathcal{M}
    \widetilde{\mathcal{M}}} \right).
  \label{balance}
\end{equation}

In this paper we shall therefore use global numerical simulations to test the
above energy balance, and to investigate the relative magnitude of the local
and and non-local transport terms. 


\section{Numerical Set-Up}
\label{setup}

\subsection{The SPH code}
All of the simulations presented hereafter were performed using a 3D smoothed
particle hydrodynamics (SPH) code, a Lagrangian hydrodynamics code capable of
modelling self-gravity (see for example, \citealt{Benz90,Monaghan92}). All
particles evolve according to individual time-steps governed by the Courant
condition, a force condition \citep{Monaghan92} and an integrator limit
\citep{BateBP95}.

We note here that with SPH, the integral of a physical quantity $A$ over a
given volume $V$ is estimated by the sum over the individual particle values
of this quantity, as below; 
\begin{equation}
  \int_{V} A\; dV \approx \sum_{{i}}
  \frac{m_{{i}}}{\rho_{{i}}} A_{{i}},
\end{equation}
where $m_{i}$ is the particle mass, and ${{i}}$ loops over all the particles
within the volume $V$.  In a similar manner, we note that a volume-averaged
value for $A$, which we shall call $\bar{A}$, can therefore be estimated via
\begin{equation}
  \bar{A} \approx \sum_{{i}}
  \frac{A_{{i}}}{\rho_{{i}}} \left/ \sum_{{i}}
  \frac{1}{\rho_{{i}}} \right. ,
  \label{Vav}
\end{equation}
when, as in all our simulations, all particles have the same mass.  

We have modelled our disc systems as a single point mass (onto which gas
particles may accrete if they enter within a  given sink radius, and satisfy
certain boundness conditions -- see \citealt{BateBP95}) orbited by 500,000 SPH
gas particles; a set up common to many other SPH simulations of such systems,
(e.g. \citealt{LodatoR04,LodatoR05,Riceetal03,Harperclark07}) but with increaesd
resolution.  The central object is free to move under the gravitational
influence of the disc.  In order to ensure the simulations were properly
converged, resolution checks were undertaken with discs consisting of both
250,000 and 1,000,000 particles -- these are discussed briefly in Appendix A.

As described in section \ref{thermo} we use a simple cooling model,
implemented in the following manner
\begin{equation} 
  \frac{du_{{i}}}{dt} = - \frac{u_{{i}}}{t_{\mathrm{cool},i}},
\end{equation}
where the $u_{{i}}$ and $t_{\mathrm{cool},i}$ are the internal energy per unit
mass and the cooling time associated with each particle respectively.  Again
as above the functional form of the cooling time is kept simple, such that
$\Omega_{{i}} t_{\mathrm{cool},i} = \beta$, where $\Omega_{{i}}$ is the
angular velocity of each particle, and where $\beta$ is held constant
throughout any particular simulation.  All simulations have been run modelling
the particles as a perfect gas, with the ratio of specific heats $\gamma =
5/3$, heat addition being allowed for via $P\mbox{d}V$ work and shock heating
and with the cooling implemented as specified above.  Artificial viscosity has
been included through the standard SPH formalism, with $\alpha_{\mathrm{SPH}}
= 0.1$ and $\beta_{\mathrm{SPH}} = 0.2$.  Note that these values are smaller
than those commonly used in SPH simulations; we use these values to limit the
transport induced by artificial viscosity. As shown in \citet{LodatoR04}, with
this choice of parameters the transport of energy and angular momentum due to
artificial viscosity is a factor of 10 smaller than that due to gravitational
perturbations, while we are still able to resolve the weak shocks occurring in
our simulations. 
  
By using the cooling prescription outlined above, the rate at which
the disc cools is governed by the dimensionless parameter $\beta$ and the
cooling is thus implemented scale free.  The governing equations of the entire
simulation can likewise be recast in dimensionless form.  In common with the
previous SPH simulations mentioned above, we define the unit mass to be that
of the central object -- the total disc mass and individual particle masses are
therefore expressed as fractions of the central object mass.  We can
self-consistently define an arbitrary unit (cylindrical) radius $R_{0}$, and
thus, with $G = 1$, the unit time is the dynamical time $t_{\mathrm{dyn}} =
\Omega^{-1}$ at radius $R = 1$. 

\begin{table}
  \begin{center}
    \begin{tabular}{cccc}
      \hline 
      $\beta$ & $q = M_{\mathrm{disc}}/M_{*}$ & No. of Particles & Duration \\
      \hline 
      4  & 0.10 & 500,000 & 4.0  $\tth$ \\
      5  & 0.10 & 500,000 & 10.0 $\tth$ \\
      6  & 0.10 & 500,000 & 10.0 $\tth$ \\
      7  & 0.10 & 500,000 & 10.0 $\tth$ \\
      8  & 0.10 & 500,000 & 10.0 $\tth$ \\
      9  & 0.10 & 500,000 & 10.0 $\tth$ \\
      10 & 0.10 & 500,000 & 10.0 $\tth$ \\
      \hline
      5  & 0.050 & 500,000 & 10.0 $\tth$ \\
      5  & 0.075 & 500,000 & 10.0 $\tth$ \\
      5  & 0.100 & 500,000 & 10.0 $\tth$ \\
      5  & 0.125 & 500,000 & 10.0 $\tth$ \\
      \hline
    \end{tabular}
    \caption{Details of numerical simulations.  Note that the duration is
      quoted in terms of the thermal time, equivalent to the cooling time at
      the outer radius $\approx 125\beta$ code units.  The $\beta = 4$ case
      fragmented, and therefore did not run for as long as the other cases.}
    \label{simulations}
  \end{center}
\end{table}

\begin{figure}
  \centering
  \includegraphics[width = 20pc]{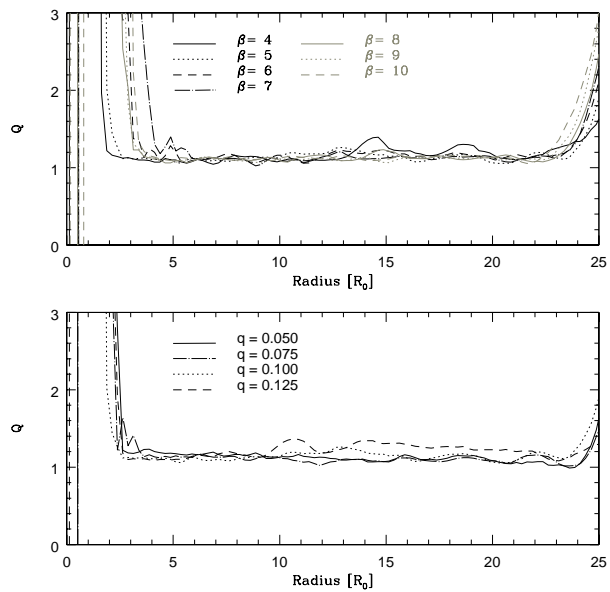}
  \caption{Profiles of $Q$ against radius for different values of the
    cooling parameter $\beta$ (top) and mass ratio $q$ (bottom) plotted at the
    times quoted in Table \ref{simulations}.}
  \label{Qprofiles}
\end{figure}

\begin{figure}
  \centering
  \includegraphics[width=20pc]{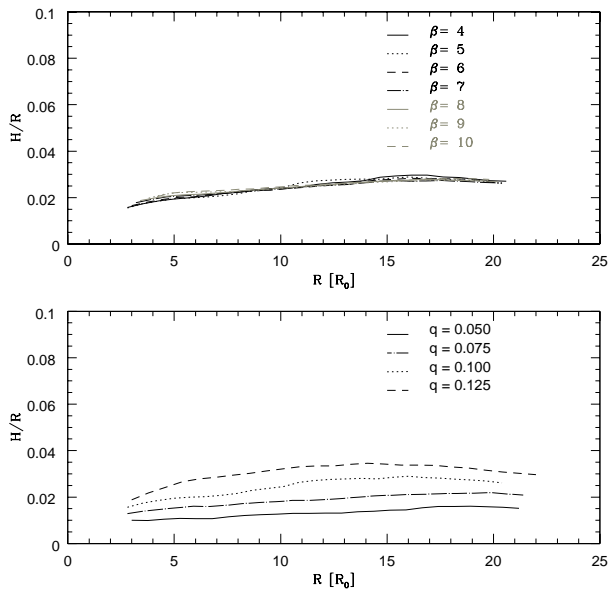}
  \caption{Disc scale height over radius ($H/R$) plotted as a function of
    radius for varying $\beta$ (top) and $q$ (bottom).  Note that in all cases
    $H/R \approx q/4$ as expected.}
  \label{HoverR}
\end{figure}

\subsection{Initial conditions}
All our simulations model a central point object of unit mass $M_{*} = 1$,
surrounded by a gaseous disc of mass $M_{\mathrm{disc}}$.  Although the bulk
of the simulations have been conducted with a disc to central object mass
ratio $q = M_{*} / M_{\mathrm{disc}}$ of 0.1, simulations have been run with
$q = 0.05, 0.075$ and $q = 0.125$ to investigate the effects of the mass ratio
on the non-local energy transport fraction $\xi$.

\begin{figure*}
  \centering
  \begin{multicols}{2}
    \includegraphics[width=20pc]{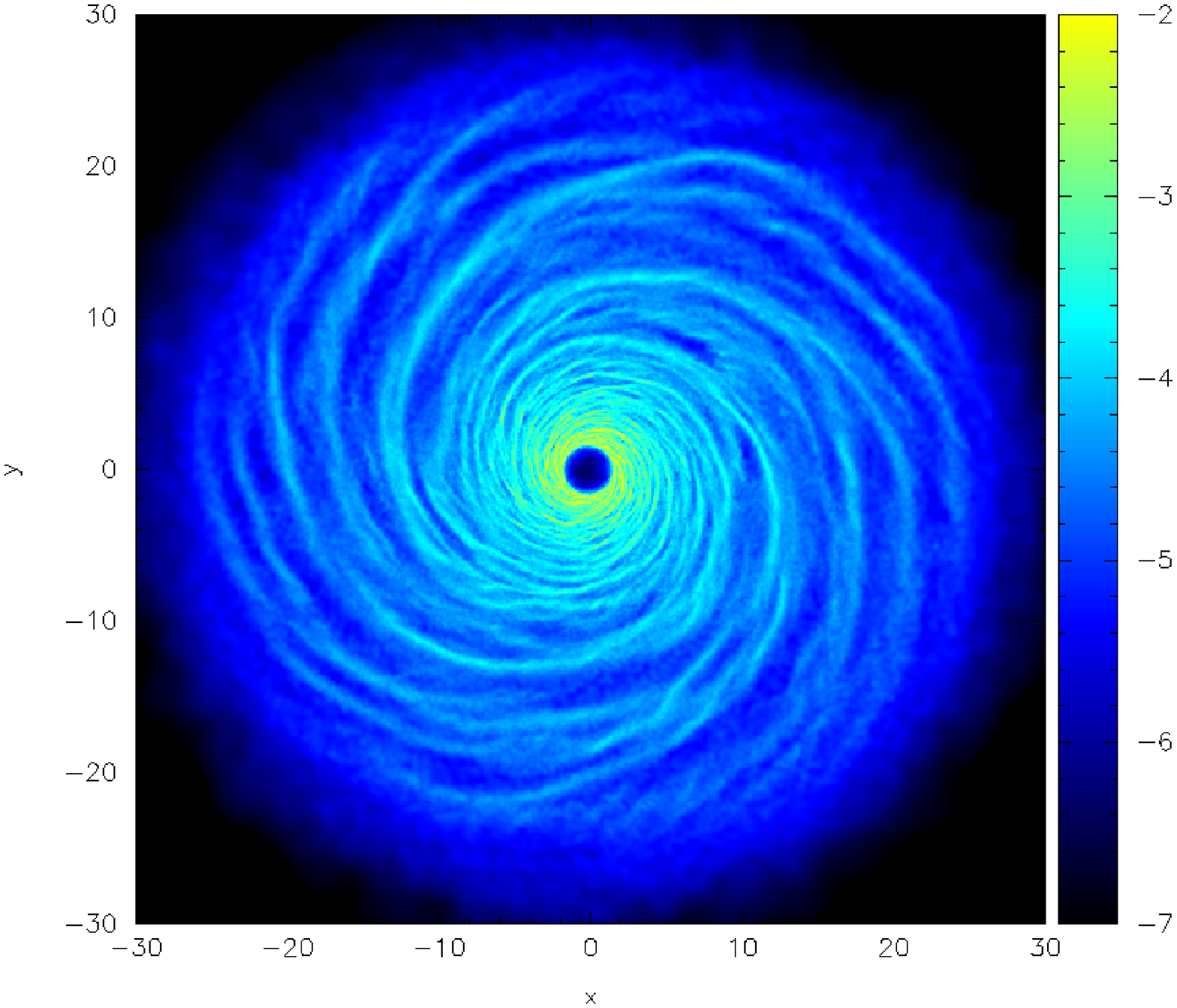}
    
    \columnbreak
    
    \includegraphics[width=20pc]{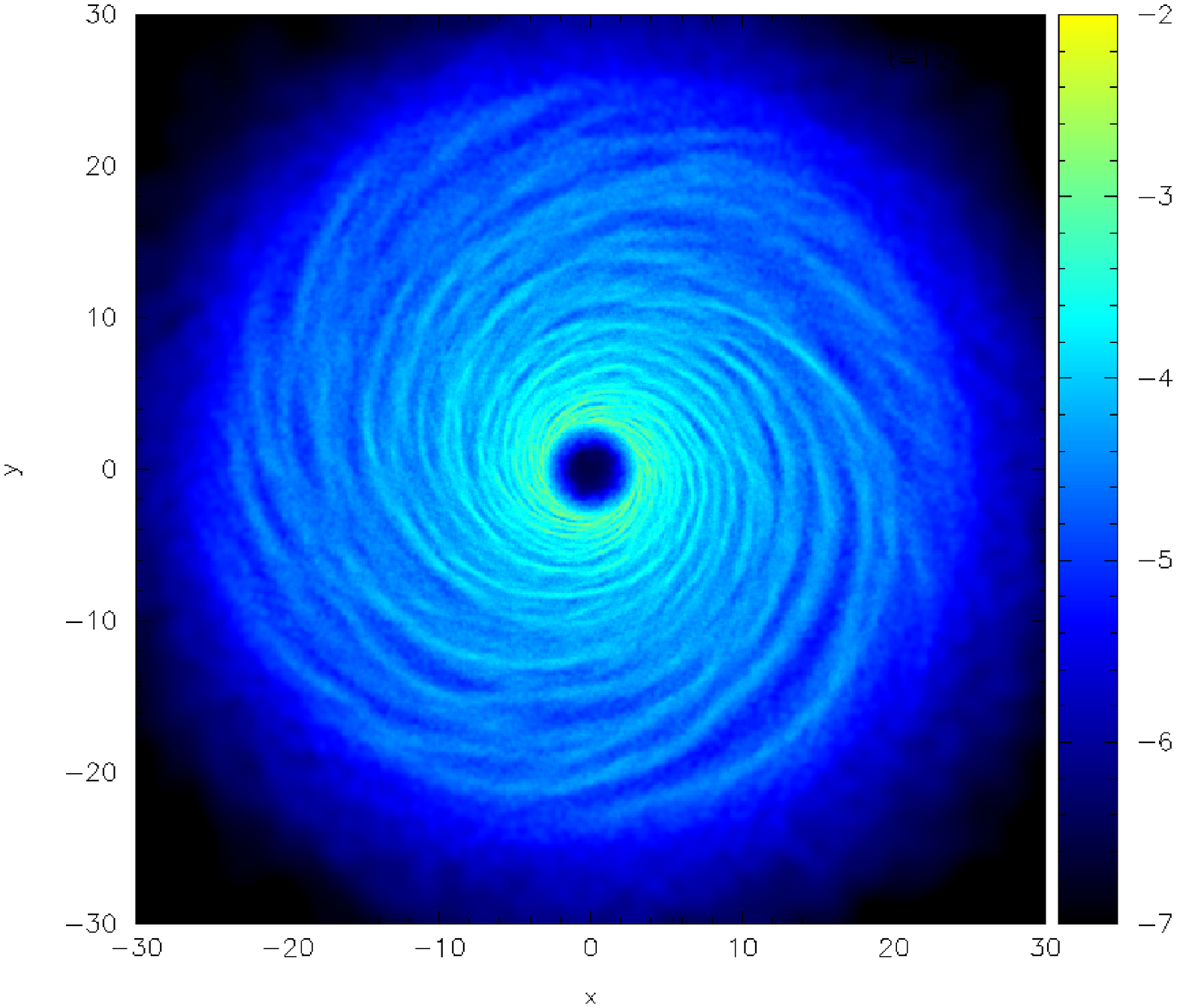}
  \end{multicols}
  \caption{Surface density structures for discs where the mass ratio $q =
    0.1$, with $\beta = 5$ (left) and $\beta = 10$ (right).  The logarithmic
    scales show surface density contours from $10^{-7}$ to $10^{-2}$ in code
    units.  Note that the direction of rotation is anticlockwise, and that the
    plots are given at the times quoted in Table \ref{simulations}.} 
  \label{rendered}
\end{figure*}

All simulations have used an initial mass surface density profile $\Sigma \sim
R^{-3/2}$, which implies that in the marginally stable state where $Q \approx
1$, the disc temperature profile should be approximately flat.  Since the
surface density evolves on the viscous time $t_{\mathrm{visc}} \gg
t_{\mathrm{dyn}} = \Omega^{-1}$ this profile remains roughly unchanged
throughout the simulations.  The initial temperature profile is
$c_{\mathrm{s}}^{2} \sim R^{-1/2}$ and is such that the minimum value of the
Toomre parameter $Q_{\mathrm{min}} = 2$ occurs at the outer edge of the disc.
In this manner the disc is initially gravitationally stable throughout.  Note
that the disc is \textit{not} initially in thermal equilibrium -- heat is not
input to the disc until gravitational instabilities are initiated. 

Radially the disc extends from $R_{\mathrm{in}} = 0.25$ to $R_{\mathrm{out}} =
25.0$, as measured in the code units described above.  The disc is initially
in approximate hydrostatic equilibrium in a Gaussian distribution of
particles with scale height $H$.  The azimuthal velocities take into
account both a pressure correction \citep{LodatoRNC07} and the enclosed disc
mass.  In both cases, any variation from dynamical equilibrium is washed out
on the dynamical timescale. 

Given the dimensions above, one outer dynamical timescale of the disc
corresponds to 125 time units.  To ensure that thermal equilibrium is reached
and that the gravitational instability is saturated, all simulations are
followed for at least 10 outer cooling times.  To this end we shall refer to
the thermal time $\tth$ for each simulation as the cooling time evaluated at
the initial outer edge of disc, taken to be at $R=25$ -- thus $\tth =
\tc(25)=\beta\Omega^{-1}(25)$.  

\subsection{Simulations run}
In all a total of nine distinct simulations were run for various values of the
cooling parameter $\beta$ and the disc to central object mass ratio $q$, as
detailed in Table \ref{simulations}.  Although previous investigations with
$\Sigma \sim R^{-1}$ have found that the fragmentation boundary is at
$\beta_{\mathrm{frag}} \approx 6$, \citep{RiceLA05,Riceetal03}, we find that
in the case where $\Sigma \sim R^{-3/2}$ the fragmentation boundary is
slightly different,  with $4 < \beta_{\mathrm{frag}} < 5$.  The simulation
where $\beta = 4$ therefore contains a fragment, and is included primarily for
completeness.  All results henceforth are given at the time quoted in the
final column of Table \ref{simulations} unless otherwise stated.  The raw data
are time-averaged over 500 unit times about these values to enhance the
signal-to-noise ratio and to give the approximate steady-state values.

\section{Simulation Results}
\label{results}
Common to all the simulations run is an initial phase in which the discs cool
rapidly until the value of $Q$ becomes approximately unity, at which point the
gravitational instability is initiated and heat is liberated to
balance the cooling.  This stage is complete after approximately one thermal
time, and from then on the discs settle into a quasi-steady state
characterised by the presence of spiral arms throughout almost their entire
radial range, with $Q \approx 1$.  The quasi-static $Q$ profiles to which the
discs converge are shown in Fig. \ref{Qprofiles}, with the cooling parameter
$\beta$ varying in the top panel, and the disc to central object mass ratio
$q$ varying in the bottom panel.  Note that the data are plotted at the times
given in Table \ref{simulations}.   Throughout all the simulations it can be
seen that the discs self-regulate to the marginally stable $Q \approx 1$
condition over a large range of radii.  

Once the disc has reached a quasi-steady state, the disc aspect ratio $H/R$
also stabilises to the value predicted by the self-regulation condition $Q
\approx 1$:
\begin{equation}
  \frac{H}{R} \approx \frac{\pi \Sigma(R) R^{2}}{M_{\star}},
\end{equation}
and is shown as a function of radius in Fig. \ref{HoverR} for different values
of $\beta$ (top panel) and $q$ (bottom panel).

\subsection{Saturation amplitude of the instability}
Once the gravitational instability has been initiated, for the case where
$\beta \leq \beta_{\mathrm{frag}}$ (i.e. where $\beta = 4$) the amplitude of
the perturbations required to balance the cooling rises to the point where
non-linear effects dominate, leading to the fragmentation of the disc into
bound objects.  In the case where $\beta > \beta_{\mathrm{frag}}$ however, the
amplitude increases on the dynamical timescale until the disc reaches dynamic
thermal equilibrium, at which point the amplitude of the surface density
fluctuations becomes constant, and the heating they provide balances the
imposed cooling.  This is observed in all our simulations where $\beta \geq 5$. 

From the simulations we can now test the prediction for the saturation
amplitude provided by Eq.(\ref{balance}) numerically.  Fig. \ref{rendered}
shows images of the surface density of the disc for the two cases $\beta=5$
and $\beta=10$, respectively, where in both cases the mass ratio is
$q=0.1$. It can be seen that, while the overall disc structure remains
essentially  constant (as confirmed by a more detailed Fourier analysis, see
below), the spiral wave amplitude as characterised by the surface density
contrasts appears to decrease with increasing $\beta$.  Noting that the
direction of rotation of the discs is anticlockwise, we find that throughout
our simulations the waves excited are all trailing waves -- they all
point in opposition to the direction of rotation.

\begin{figure}
  \centering
  \includegraphics[width=20pc]{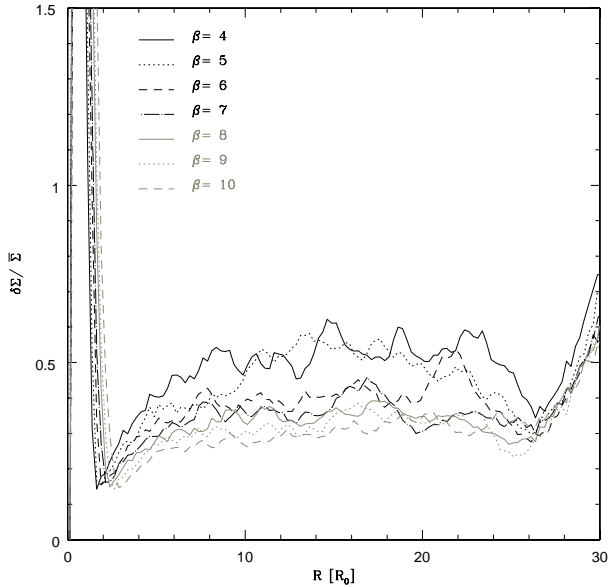}
  \caption{Variation of the relative mass surface density perturbation
    amplitude $\dSoS$ with radius for various values of the cooling
    parameter $\beta$.  All data plotted at the times shown in Table
    \ref{simulations}.}
  \label{dSoS_beta}
\end{figure}

While the SPH code allows us to conduct a global 3D simulation of discs with
relative ease, it does not permit the direct calculation of an intrinsically
2D quantity, such as the surface density perturbation amplitude $\delta \Sigma
/ \Sigma$.  Therefore, in order to calculate this quantity, we overlay a
cylindrical grid on the disc, such that each cell contains approximately
$N_{\mathrm{neigh}}$ particles, where $N_{\mathrm{neigh}} \approx 50$ is the
average number of neighbours within a smoothing kernel for our
simulations. For each annulus of cells we can calculate an average surface
density $\bar{\Sigma}$, and by comparing this to the value calculated for each
cell within this annulus we can evaluate an annulus averaged RMS value for the
perturbation amplitude $\delta \Sigma / \bar{\Sigma}$.  This is shown as a
function of radius $R$ and the cooling parameter $\beta$ in
Fig. \ref{dSoS_beta}.

\begin{figure}
  \centering
  \includegraphics[width=20pc]{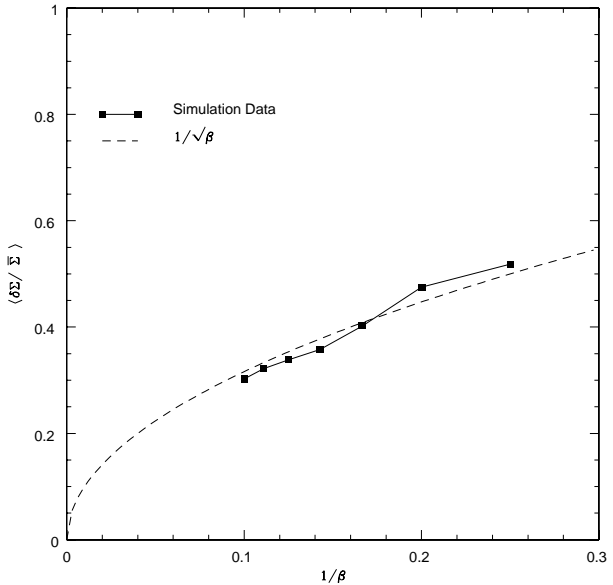}
  \caption{Variation of the radially and azimuthally averaged relative
    surface density perturbation amplitude $\dSoS$ with the inverse cooling
    parameter $1/\beta$.  The radial average is calculated over the range $5
    \leq R \leq 24$.}
  \label{dSoS_average}
\end{figure}

\begin{figure}
  \centering
  \includegraphics[width=20pc]{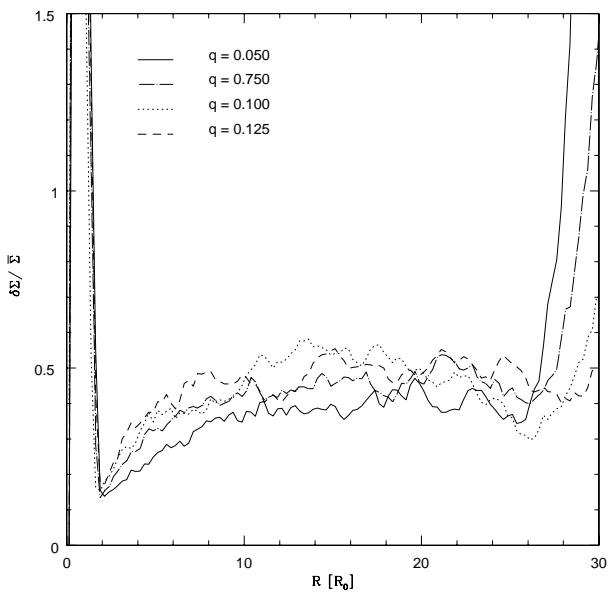}
  \caption{Variation of the relative mass surface density perturbation
    amplitude $\dSoS$ with radius for various values of the disc to central
    object mass ratio $q$.}
  \label{dSoS_q}
\end{figure}

\begin{figure*}
  \centerline{
    \epsfig{figure=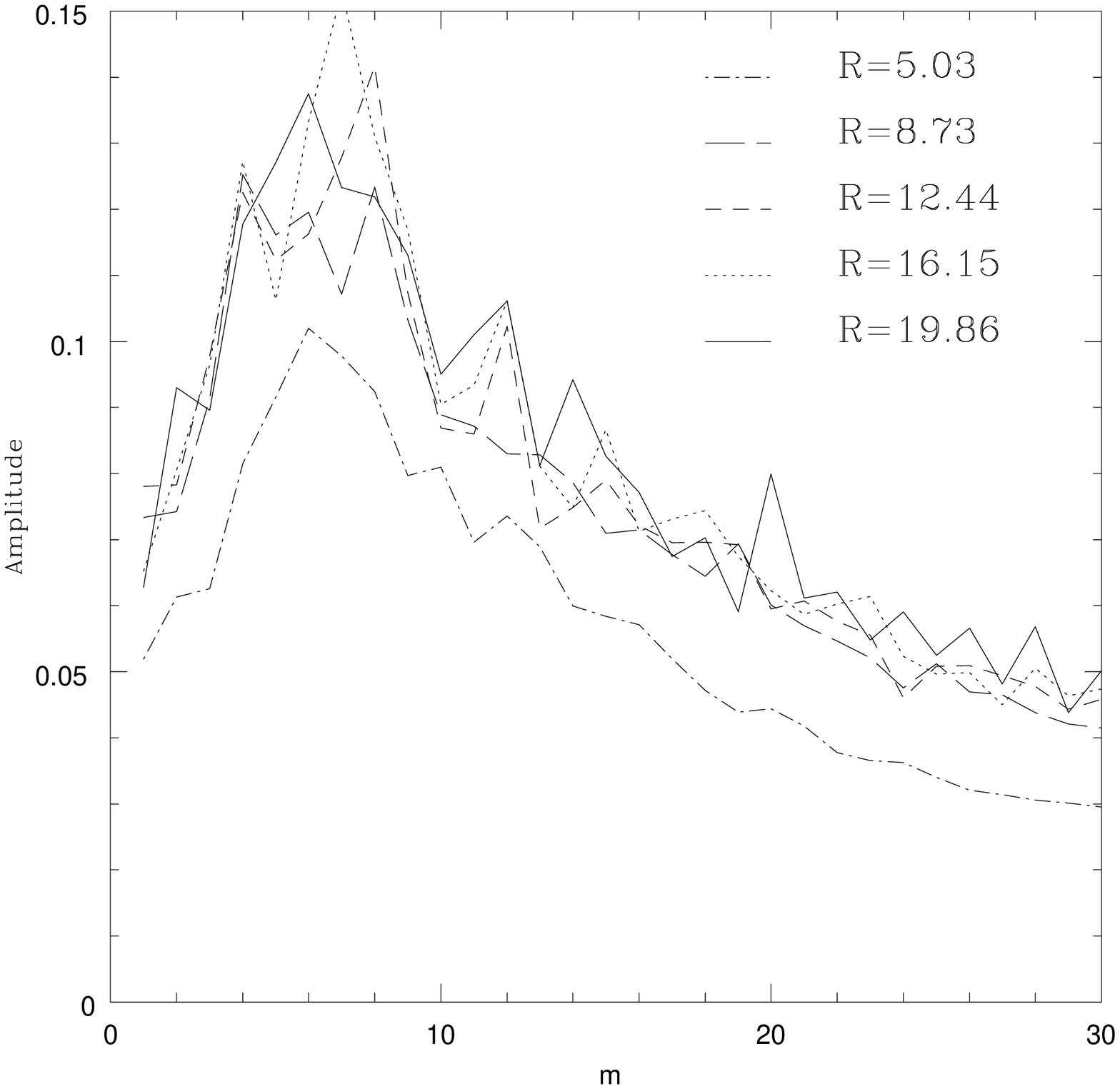,width=0.3\textwidth}
    \epsfig{figure=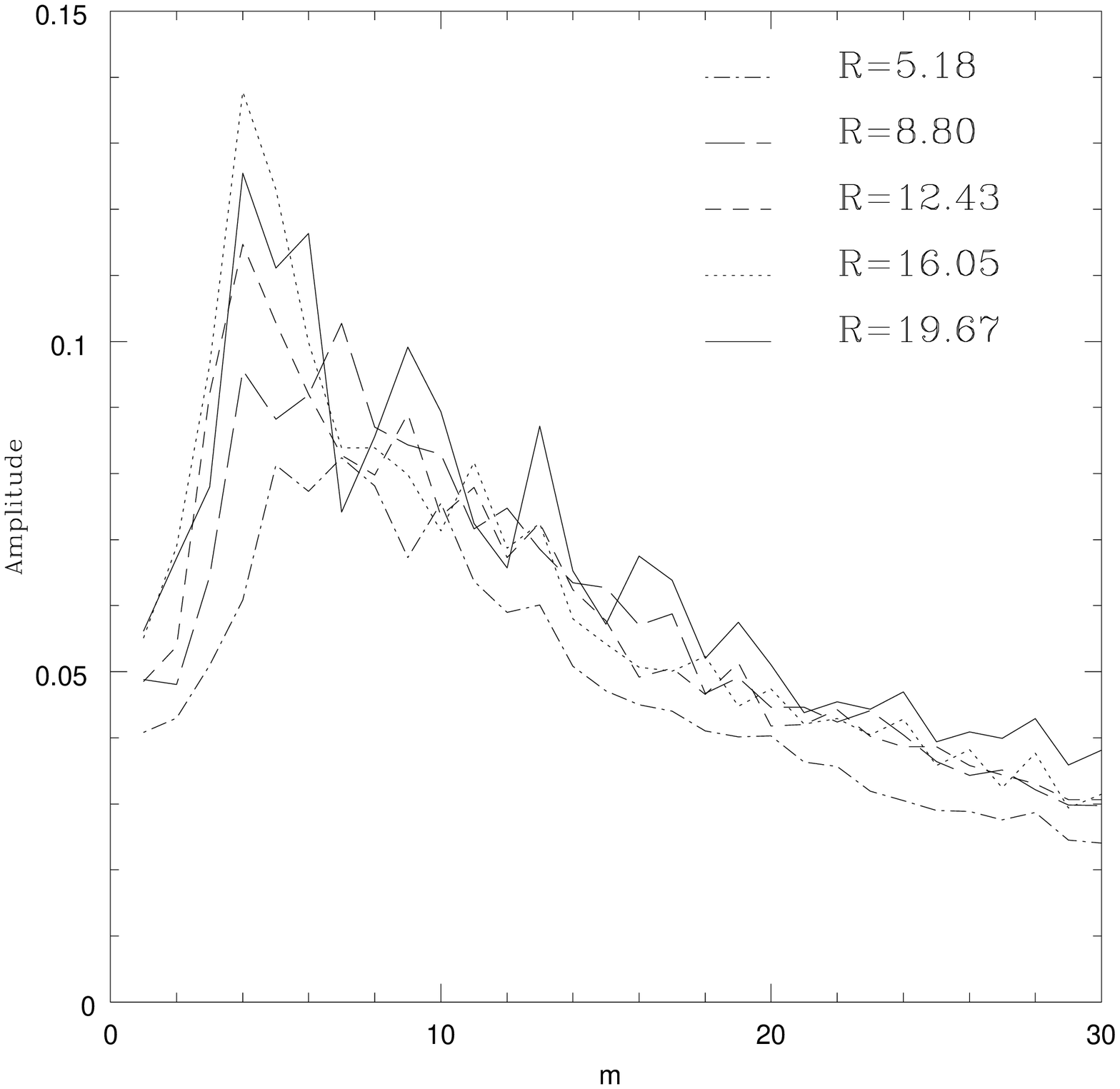,width=0.3\textwidth}
  }
  \centerline{
    \epsfig{figure=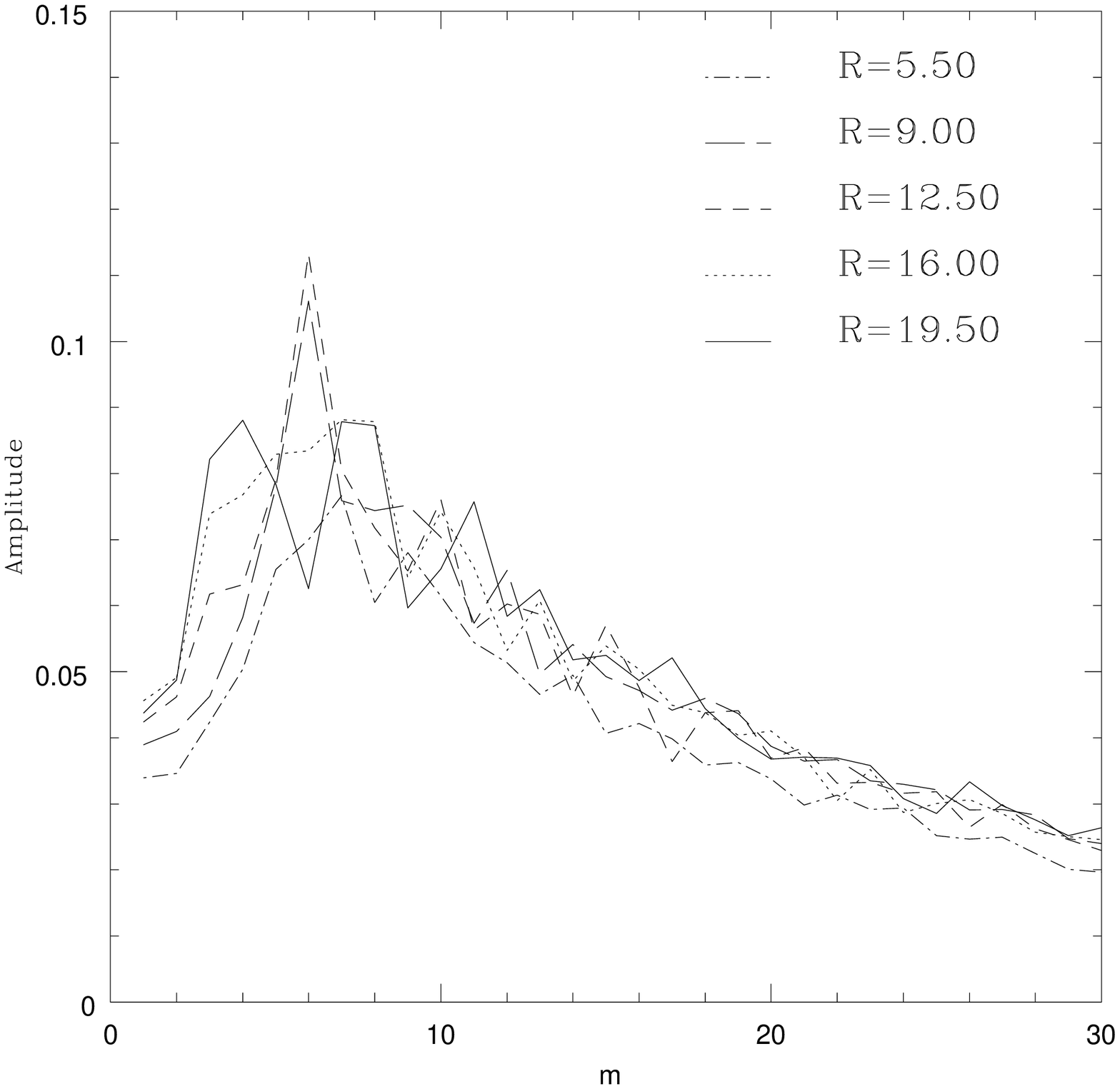,width=0.3\textwidth}
    \epsfig{figure=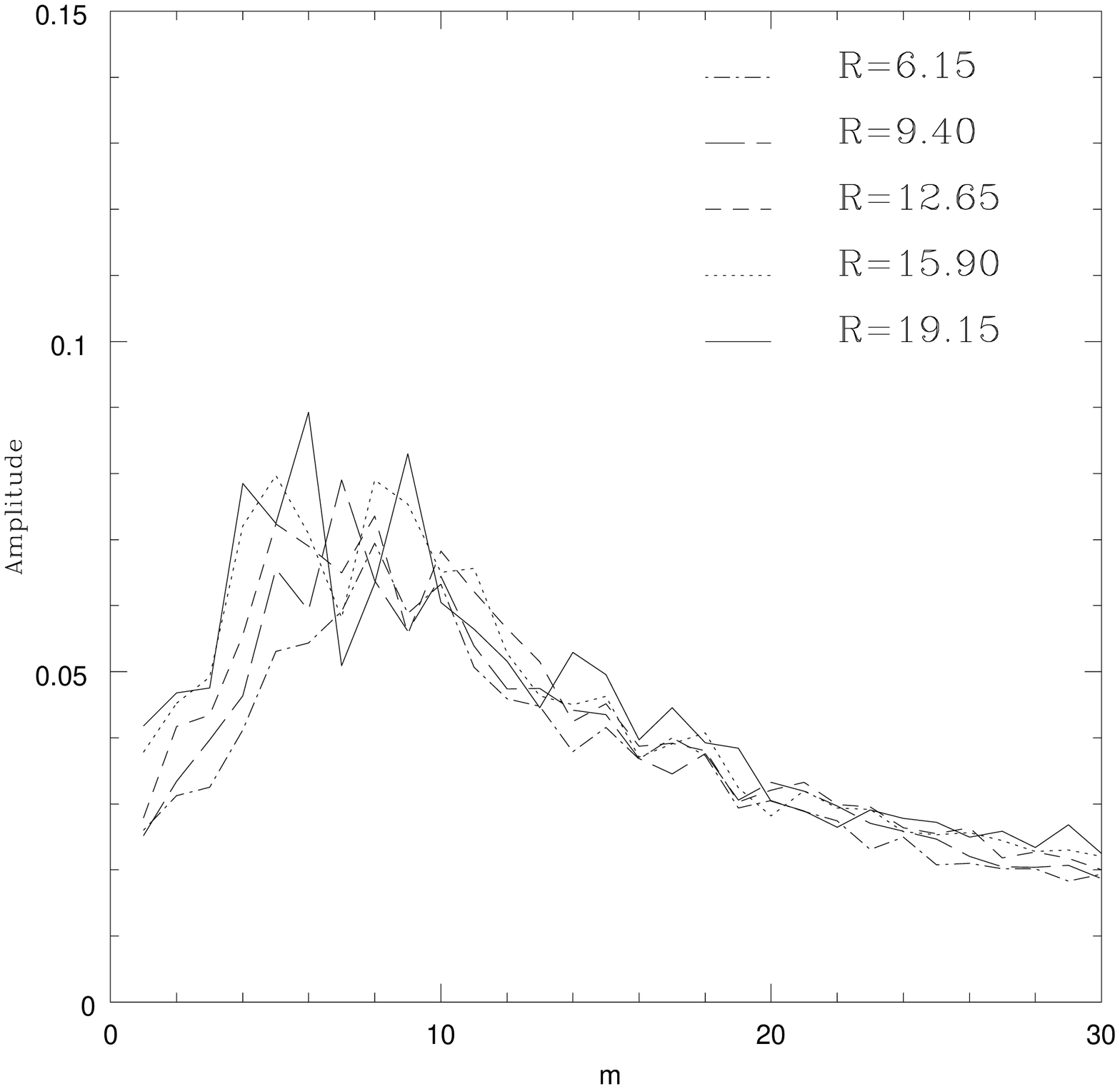,width=0.3\textwidth}
  }
  \caption{Azimuthal mode amplitudes excited at various radii where $\beta =
    4$ (top left), $\beta = 6$ (top right), $\beta = 8$ (bottom left) and
    $\beta = 10$ (bottom right).}
  \label{mmodes_amplitudes_beta}
\end{figure*}

From Fig. \ref{dSoS_beta}, it is clear that there is an increasing trend in
$\dSoS$ with decreasing $\beta$ and that furthermore, away from the disc
boundaries the saturation amplitude is approximately constant with radius.
The low values for the perturbation amplitude at small radii ($R \lesssim 5$)
are probably due to the increased number of particles per grid cell smoothing
out the underlying variation.  We can however characterise the strength of
the perturbation by simply averaging $\dSoS$ over the self-regulated portion
of the disc, that we define as $5\leq R \leq 25$ (cf. Fig
\ref{Qprofiles}). Fig. \ref{dSoS_average} shows the relation between the
azimuthally and radially averaged amplitude, which we denote as
$\langle\dSoS\rangle$, and the cooling parameter $\beta$. Each point
represents a single simulation, while the curve shows our best fit to the data
using the inverse square root dependence predicted by Eq.(\ref{balance}). From
the simulations we therefore obtain the following empirical relationship, for
the case where $q = 0.1$ 
\begin{equation} 
  \left \langle \frac{\delta\Sigma}{\bar{\Sigma}}\right\rangle \approx
  \frac{1.0}{\sqrt\beta}. 
  \label{empiricaldSoS}
\end{equation}

\begin{figure}
  \centering
  \includegraphics[width=20pc]{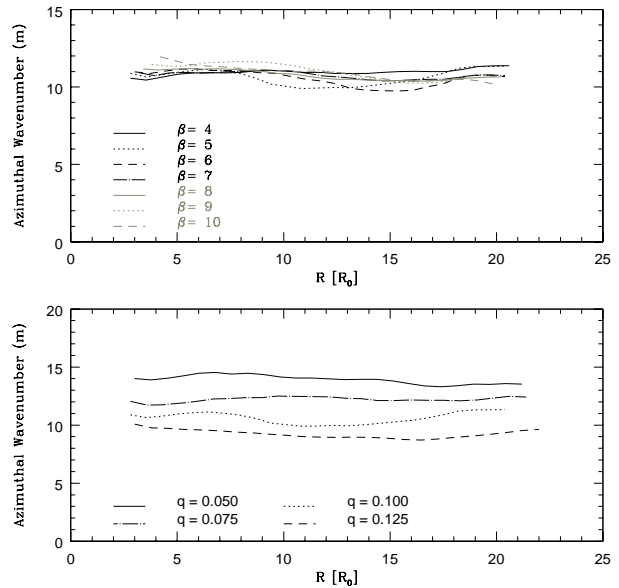}
  \caption{Variation of the average azimuthal wavenumber excited as a
    function of wavenumber for $\beta = 4$ -- $10$ where $q=0.1$ (top) and as
    $q$ varies with $\beta = 5$ (bottom).}
  \label{mmodeaverages}
\end{figure}

In a similar manner we can also calculate the variation in $\dSoS$ with $q$,
which is shown in Fig. \ref{dSoS_q}.  We see that the strength of the
perturbation tends to increase with the mass ratio $q$, although it is clear
that this dependence on $q$ is rather less than linear. 

\subsection{Fourier analysis: azimuthal structure}
From the simulations we have found empirically that the perturbation strength
$\left \langle \delta \Sigma / \bar{\Sigma} \right \rangle$ follows a
$\beta^{-1/2}$ relationship, as predicted by Eq.(\ref{balance}).  However,
this equation also shows a clear dependence on the wave modes excited within
the disc through the action of the gravitational instability.  To elucidate
this relationship further we have therefore conducted a Fourier analysis of
the wave modes in the disc, a full description of which may be found in
Appendix B.  In this section we therefore consider the effects of both the
cooling (via $\beta$) and the disc to central object mass ratio $q$ on the
excitation of the azimuthal $m$ wavenumbers -- the next section will describe
the excitation of the radial wavenumbers. 

\begin{figure*}
  \centering
  \begin{multicols}{2}
    \includegraphics[width=20pc]{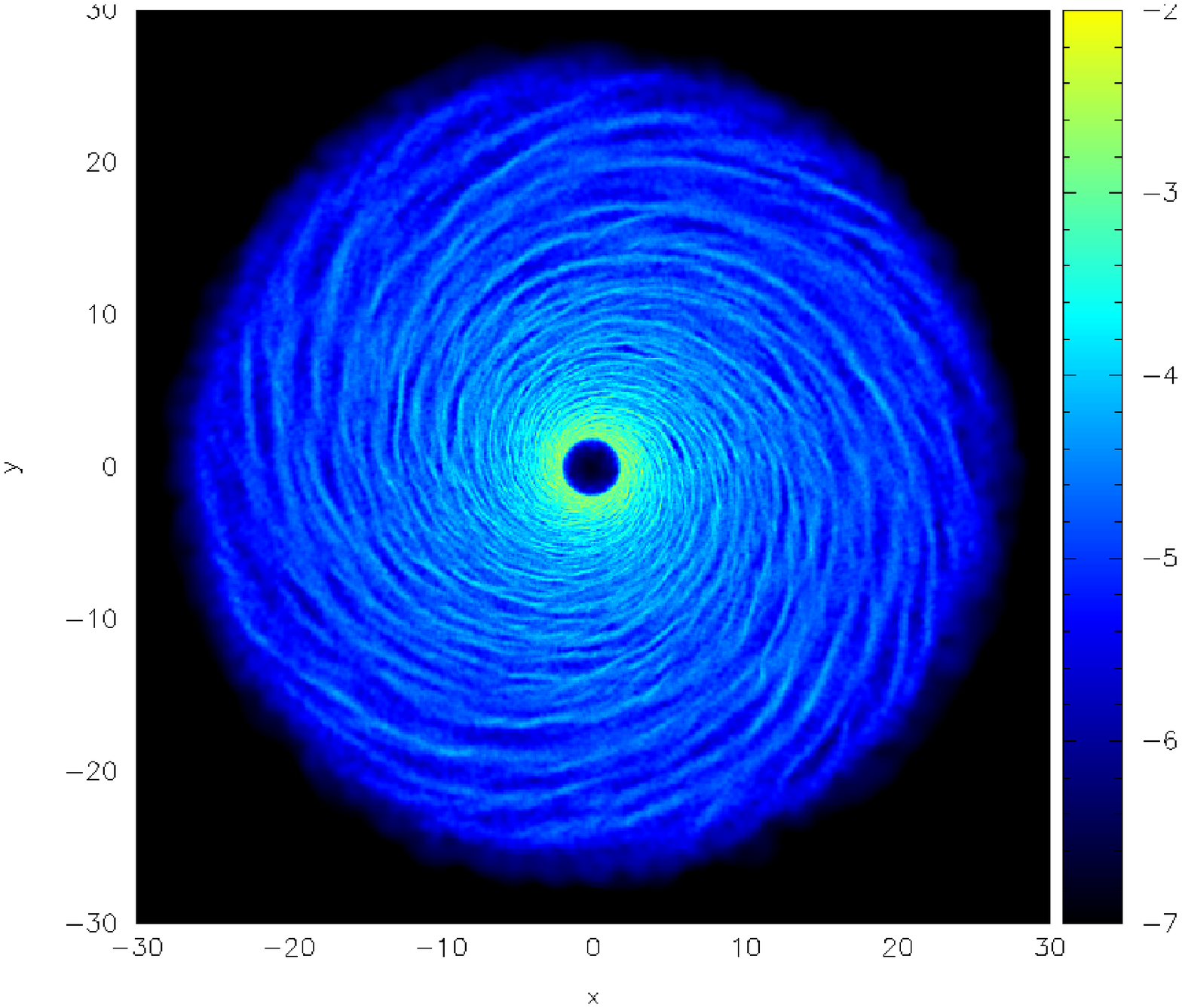}
    
    \columnbreak
    
    \includegraphics[width=20pc]{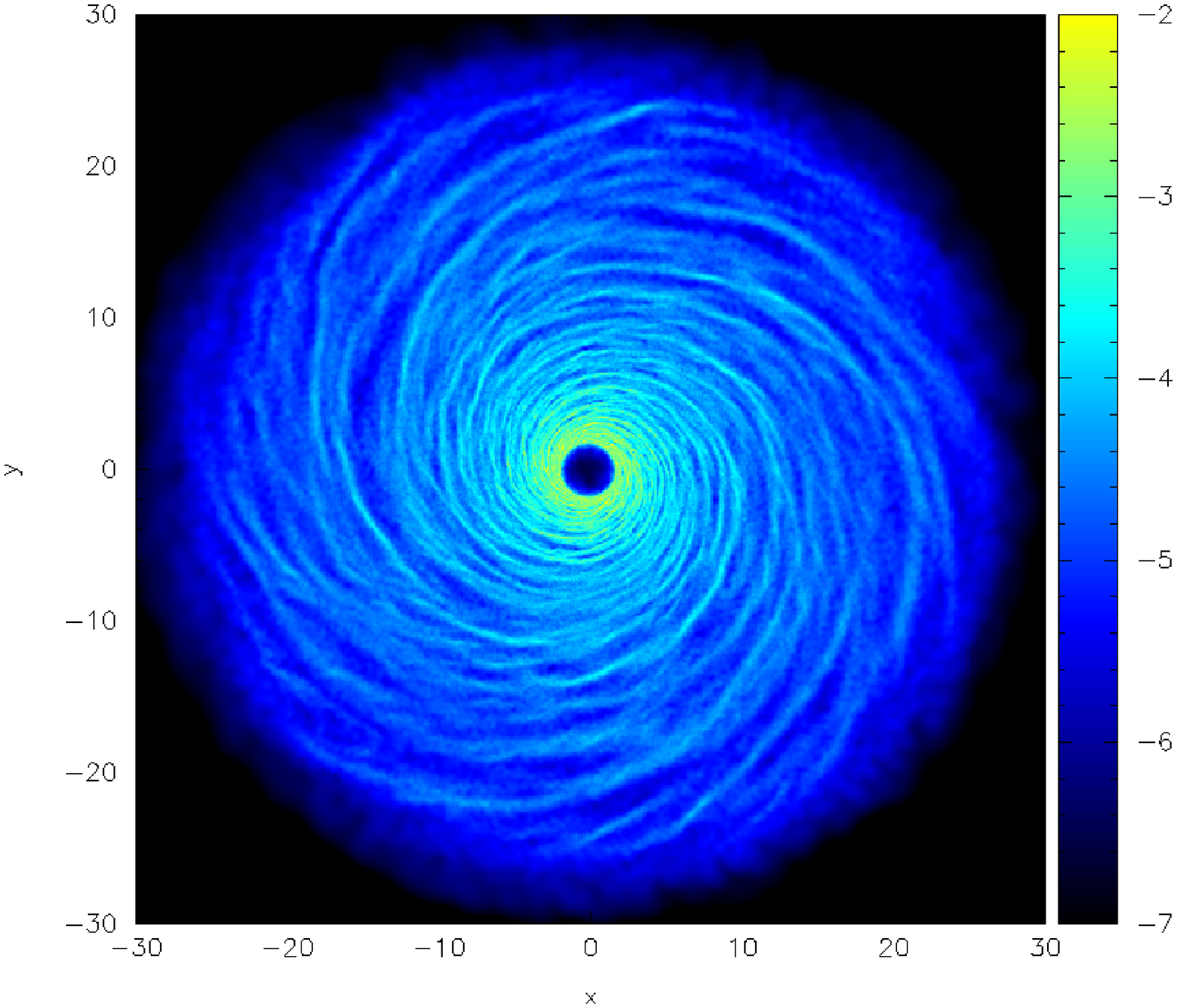}
  \end{multicols}
  \caption{Surface density structures for discs with $q = 0.05$ (left) and
    $q= 0.125$ (right), with $\beta = 5$.  The logarithmic scale shows mass
    surface density contours from $10^{-7}$ to $10^{-2}$ in code units.
    These therefore form a direct comparison with Fig. \ref{rendered},
    where $q = 0.1, \; \beta= 5$.  Once again the direction of rotation of
    the discs is anticlockwise.} 
  \label{renderedq}
\end{figure*}

\begin{figure*}
  \centerline{
    \epsfig{figure=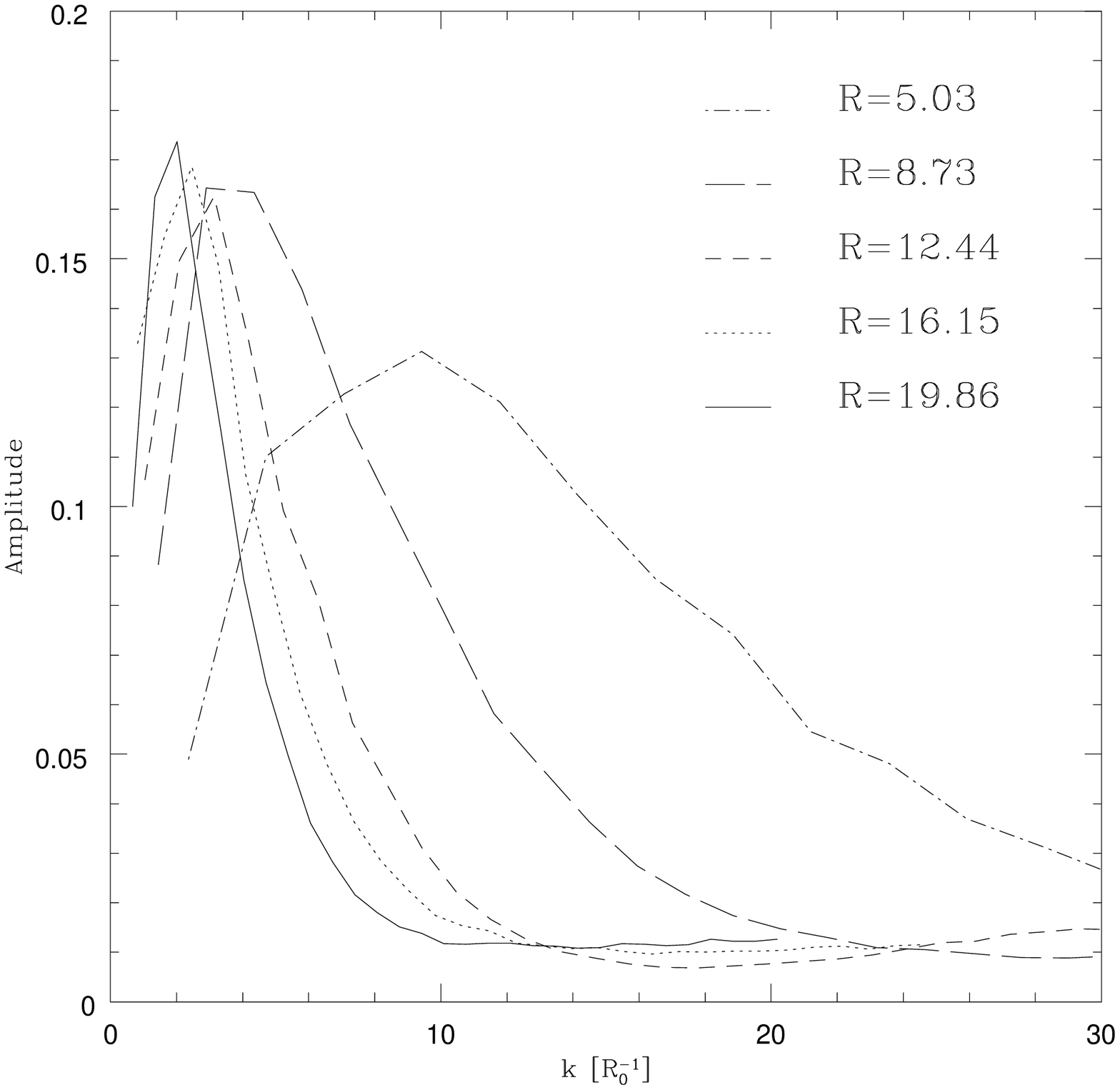,width=0.3\textwidth}
    \epsfig{figure=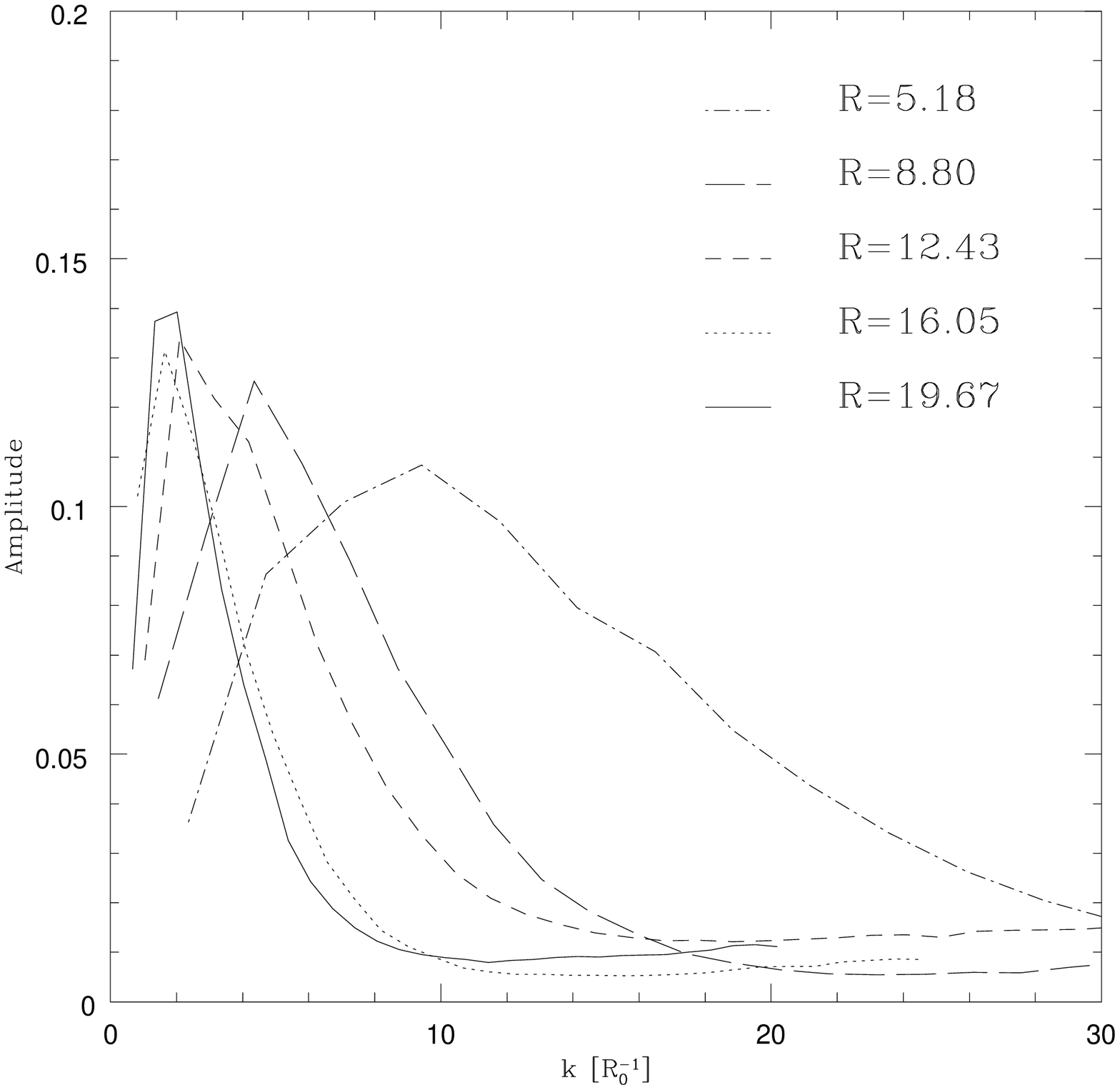,width=0.3\textwidth}
  }
  \centerline{
    \epsfig{figure=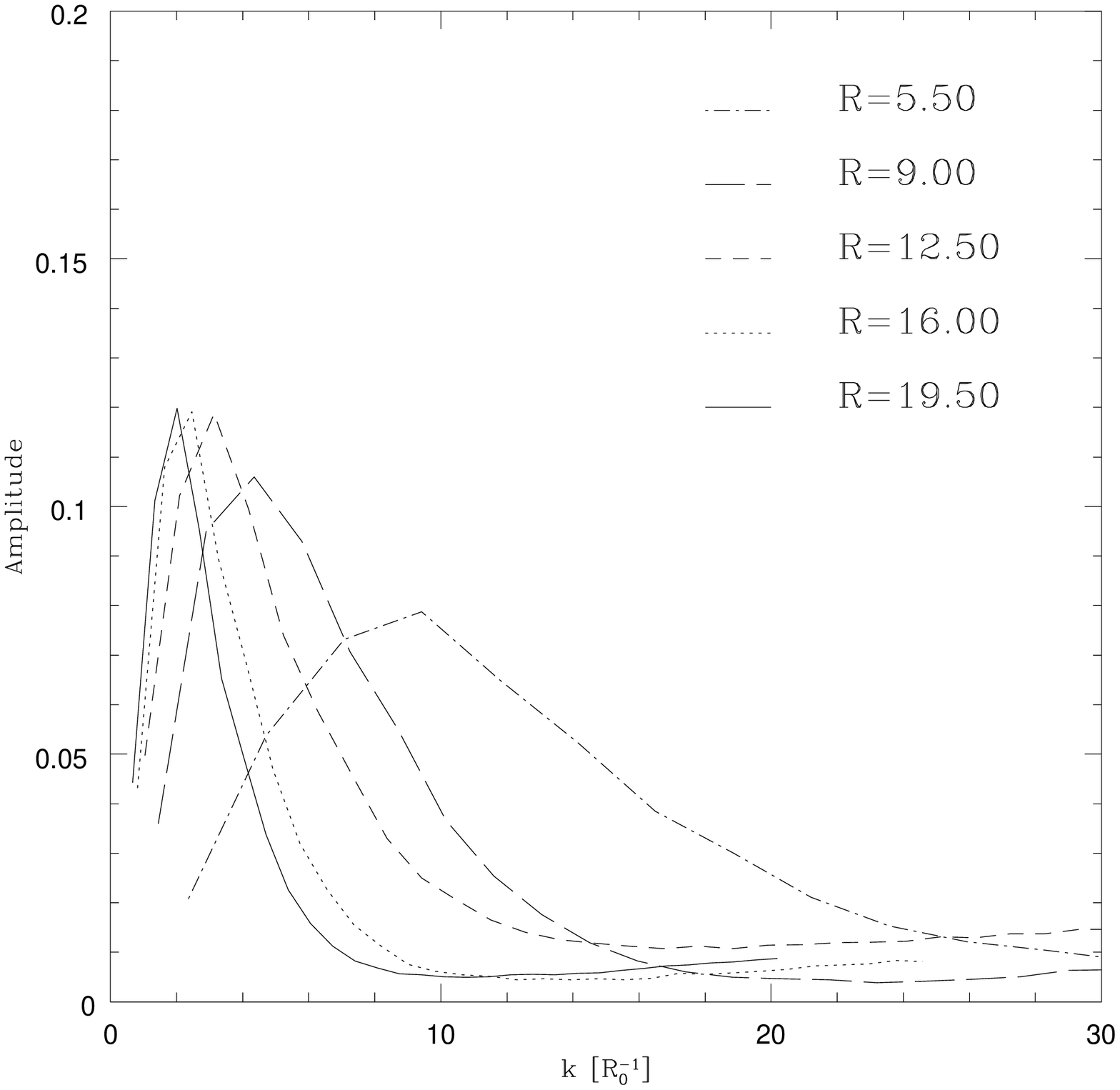,width=0.3\textwidth}
    \epsfig{figure=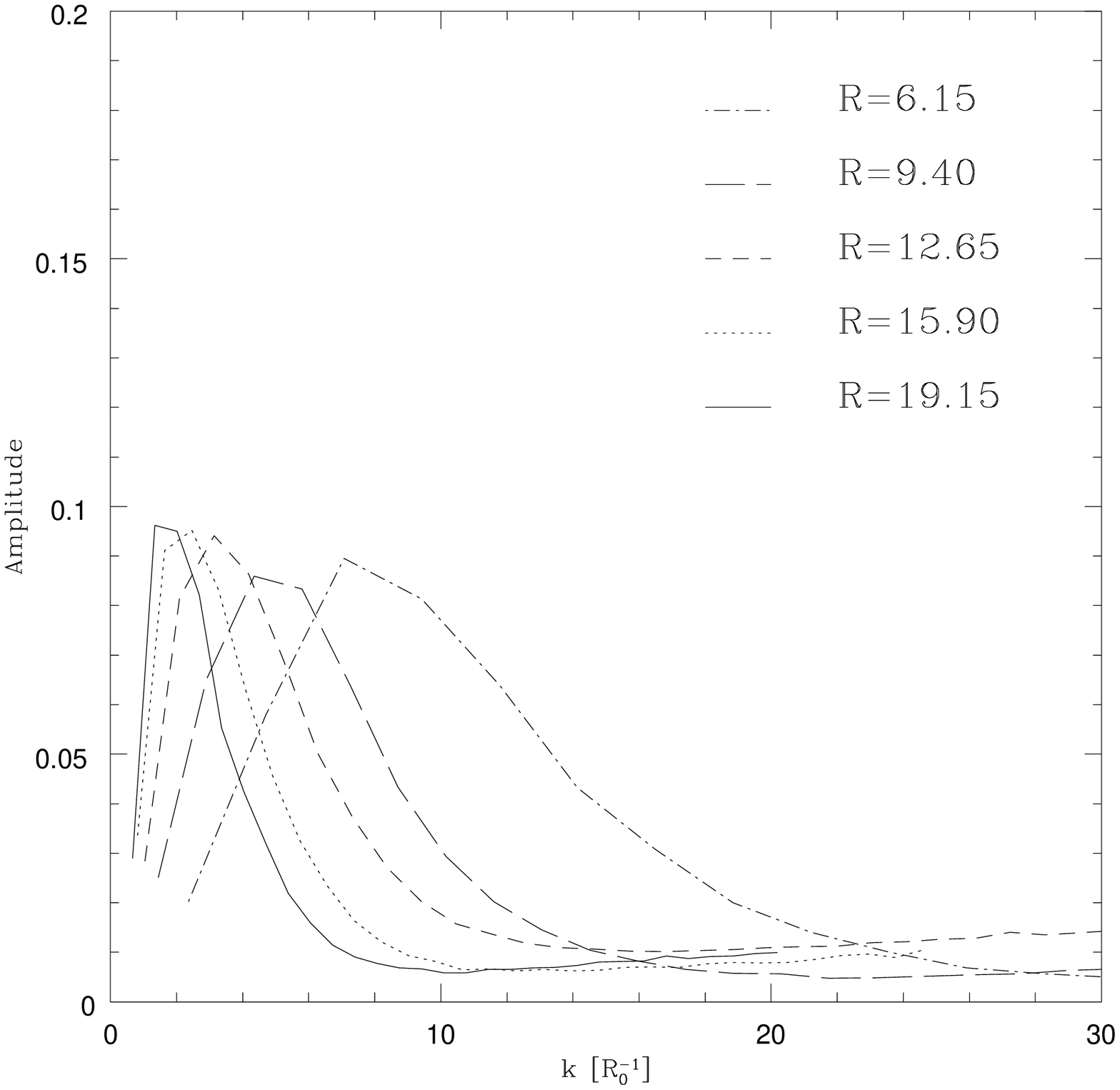,width=0.3\textwidth}
  }
  \caption{Radial mode amplitudes excited at various radii where $\beta =
    4$ (top left), $\beta = 6$ (top right), $\beta = 8$ (bottom left) and
    $\beta = 10$ (bottom right).} 
  \label{kmodes_amplitudes_beta}
\end{figure*}

\begin{figure*}
  \centerline{
    \epsfig{figure=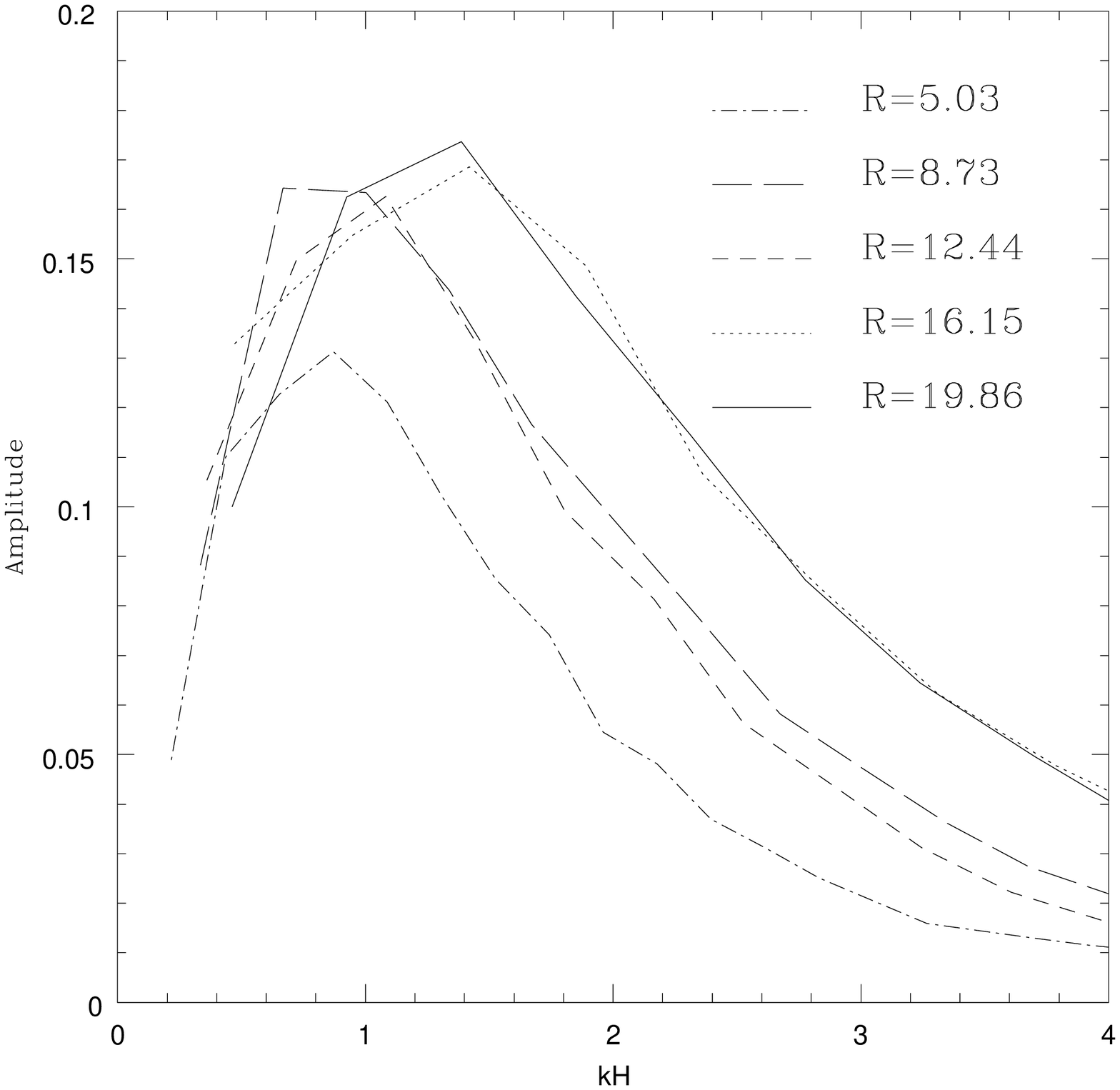,width=0.3\textwidth}
    \epsfig{figure=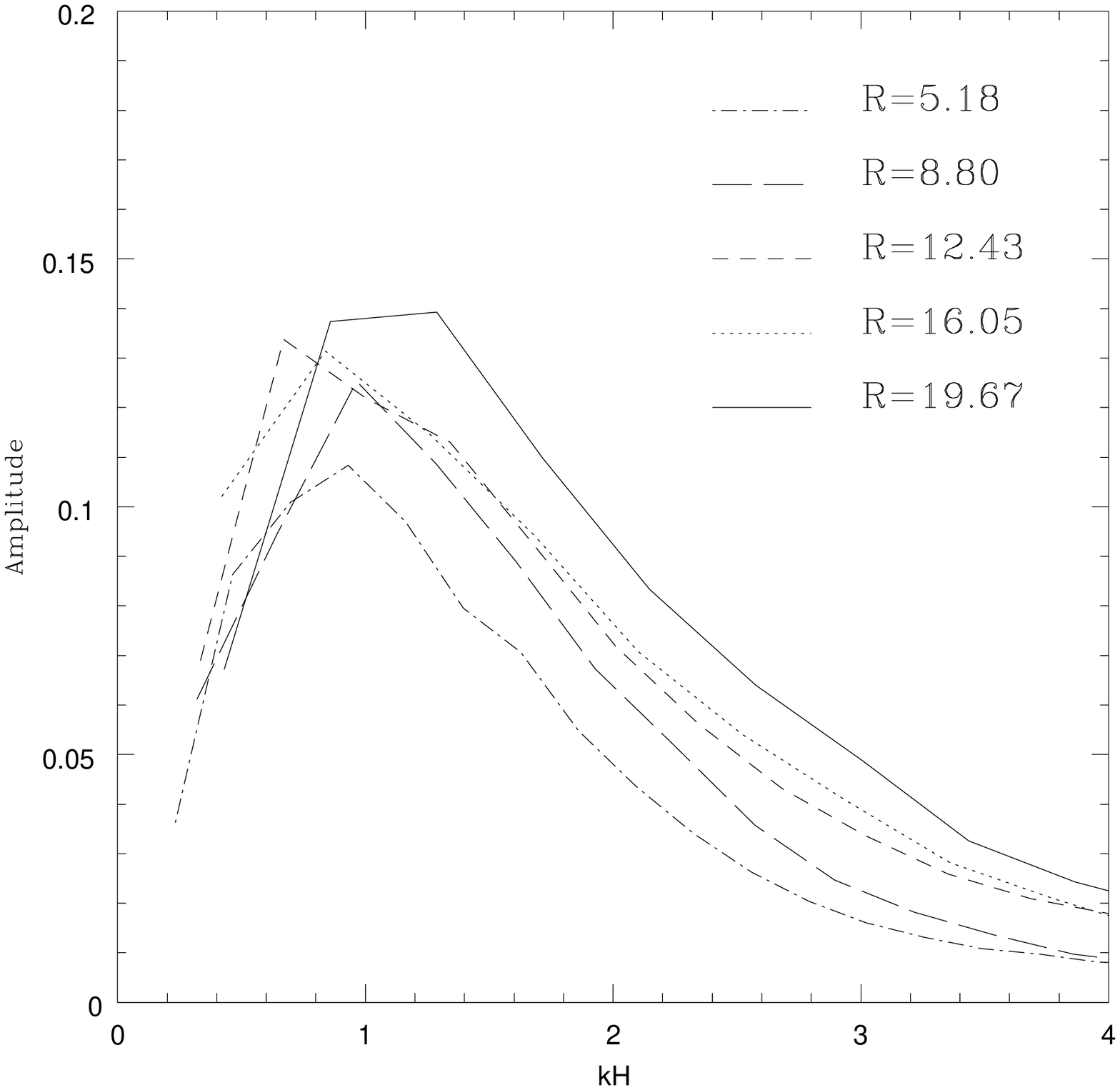,width=0.3\textwidth}
  }
  \centerline{
    \epsfig{figure=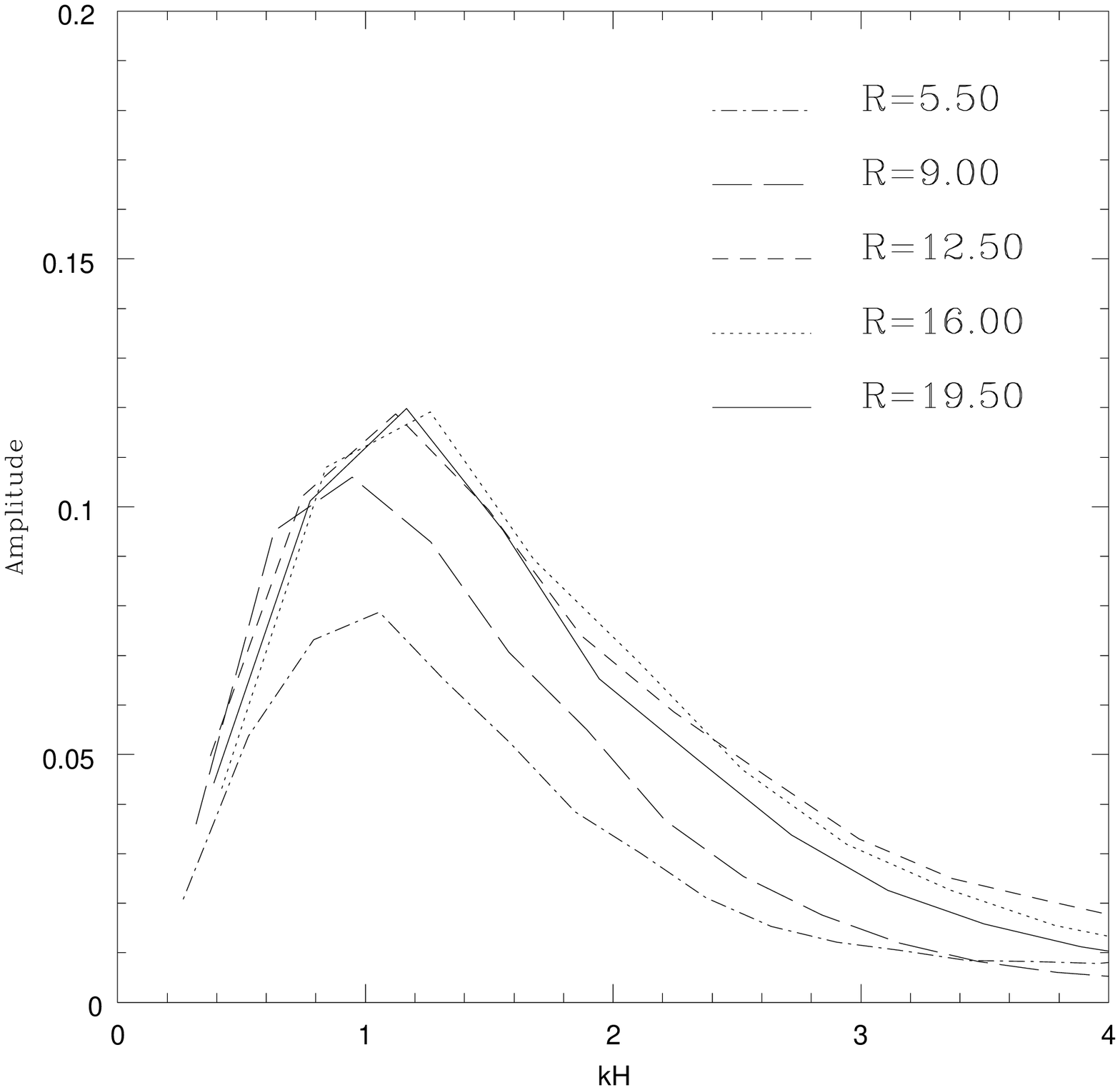,width=0.3\textwidth}
    \epsfig{figure=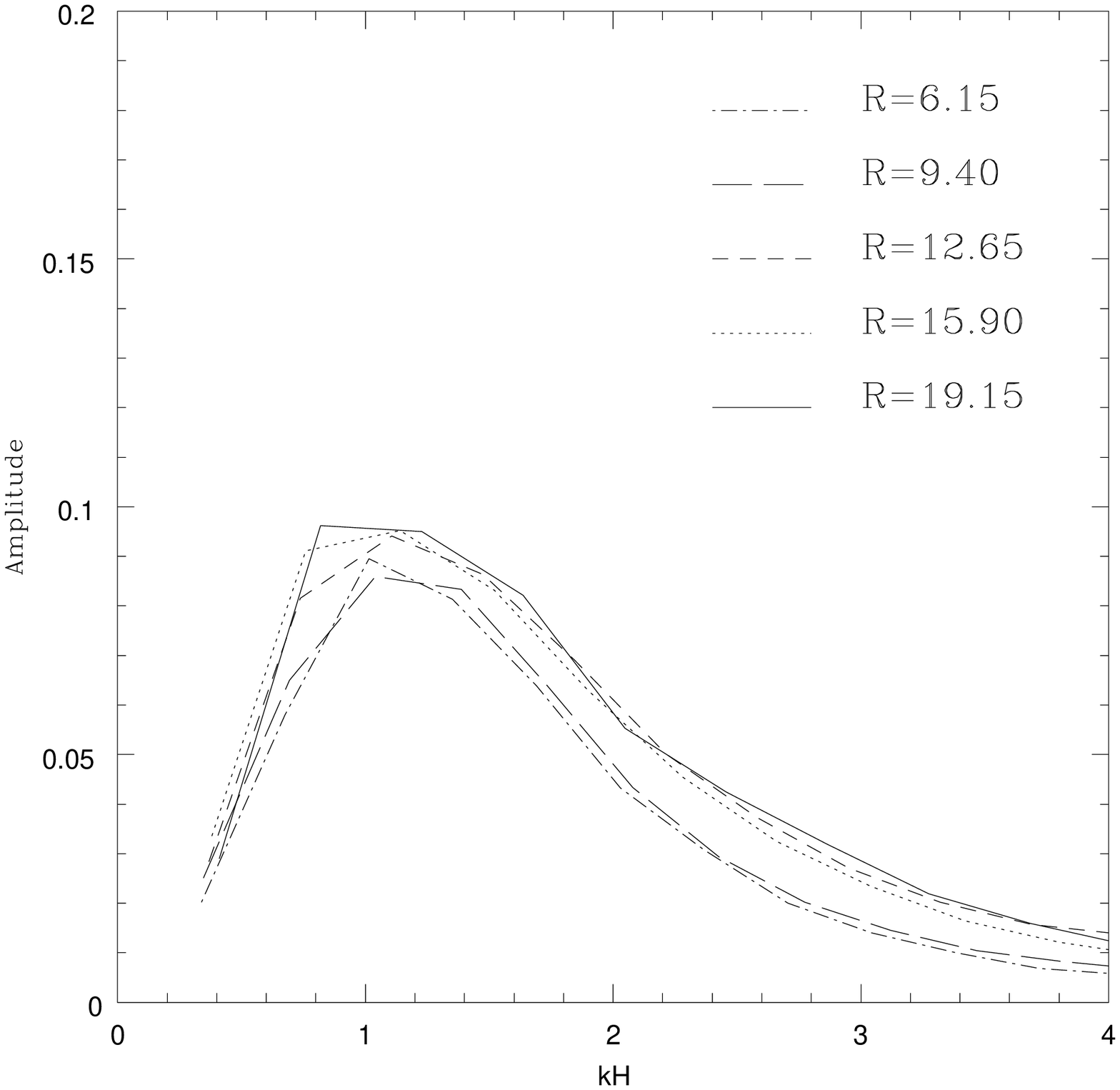,width=0.3\textwidth}
  }
  \caption{Mode amplitudes plotted against the product of the peak radial
    modenumber and the disc scale height $H$ excited for various radii where
    $\beta = 4$ (top left), $\beta = 6$ (top right), $\beta = 8$ (bottom
    left) and $\beta = 10$ (bottom right).} 
  \label{kmodes_kH_beta}
\end{figure*}

We find that in general, whatever the imposed cooling regime, for a given mass
ratio $q = 0.1$ the distribution of the azimuthal wavenumbers determined by
the gravitational instability remains approximately constant, with the
dominant mode at around $m \approx 5$.  The mode distributions at five radii
throughout the disc for the cases where $\beta = 4,6,8$ and $10$ are shown in
Fig. \ref{mmodes_amplitudes_beta}.  It is clear that the spectral distribution
of the modes shows little variation with radius and with the imposed cooling,
except that the amplitude of the modes decreases as $\beta$ increases, as
expected in view of the decreased perturbation amplitudes seen in Figs
\ref{rendered}, \ref{dSoS_beta} and \ref{dSoS_average}. Additionally, Fig.
\ref{mmodeaverages} (top panel) shows the variation of the average
wavenumber against radius for all the values of $\beta$ that have been tested,
and we note that although some small variation is seen, it is uncorrelated
with the imposed cooling.

The bottom panel of Fig. \ref{mmodeaverages} shows the variation in the
average azimuthal modes excited as the disc to central object mass ratio $q$
varies, while the cooling is held constant with $\beta = 5$.  It can be seen
from the plot that variation of this parameter does have a marked effect on
the power spectrum of the waves -- the average mode number varies inversely
with the mass ratio, from $m_{\mathrm{av}} \approx 15$ where $q = 0.05$ to
$m_{\mathrm{av}} \approx 10$ where $q = 0.125$.  This variation is also clearly
seen in Fig. \ref{renderedq}, where a large number of flocculent arms are
present in the disc with $q = 0.05$, and fewer, rather more well-defined
spiral arms appear in the disc where $q = 0.125$ (cf. the similar result
obtained in \citealt{LodatoR04}).  By comparison, the left panel of
Fig. \ref{rendered} shows a disc with $\beta = 5$ and $q = 0.1$, and the
pattern of spiral arms present is intermediate to those shown in
Fig. \ref{renderedq}.   

\subsection{Fourier analysis: radial structure}
We now consider the radial wavenumbers $k$ of the waves excited by the
gravitational instability. In contrast to the azimuthal modes, it is clear
from Figs. \ref{rendered} and \ref{renderedq} that there is significant
variation in the radial wavenumber $k$ with radius, and Fig. \ref{renderedq}
suggests that there is an additional variation with the disc to central object
mass ratio.

In Fig. \ref{kmodes_amplitudes_beta} we show the variation in the power
spectrum of different radial wavenumbers for the cases where $\beta = 4,6,8$
and $10$ and $q = 0.1$, at the same radii as the azimuthal wavenumbers shown
in Fig. \ref{mmodes_amplitudes_beta}.  As with the azimuthal modes, we see
little overall change in the spectral distribution of the modes with varying
$\beta$ excepting that the amplitudes of the modes decrease as the cooling
weakens.  Conversely however, these plots show a significant variation with
radius, in that the peak wavenumber decreases with increasing radius, and thus
the dominant wavelength similarly increases with radius.

\begin{figure}
  \centering
  \includegraphics[width=20pc]{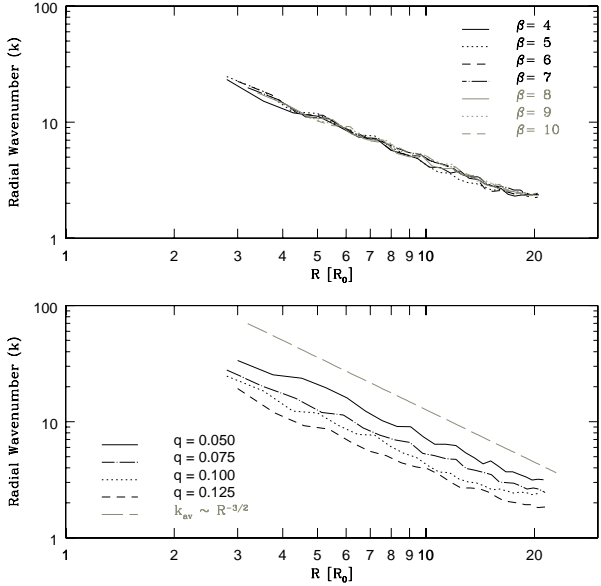}
  \caption{Variation of the average radial wavenumber as a function of
    radius for $\beta = 4$ -- $10$ and $q = 0.1$ (top) and as $q$ varies with
    $\beta = 5$ (bottom).}
  \label{kmodeaverages}
\end{figure}

Figure \ref{kmodes_kH_beta} on the other hand shows the power spectrum as a
function of $kH$, where we normalize the wavenumber to the expected most
unstable one, $H^{-1}$ (see Section 2.1). We thus confirm the expectations from
the linear WKB approach, as these plots show a clear peak at $kH\approx
1$. This therefore suggests that throughout the disc the excited waves are
close to co-rotation. Fig. \ref{kmodeaverages} (top panel) shows the average
radial wavenumber as a function of radius for all the values of $\beta$
considered, again confirming the trends already discussed and also further
showing that, excepting the variation in amplitude discussed above, the
structure excited by the simulation is essentially independent of the cooling
imposed.  It also shows that the simulation to simulation scatter is very
small.    

The bottom panel of Fig. \ref{kmodeaverages} shows that, as with the azimuthal
wavenumbers, there is clear variation in the average radial wavenumber
$k_{\mathrm{av}}$ with the disc to central object mass ratio for a given
$\beta$ (in this case $\beta=5$); increasing the mass ratio decreases the
average wavenumber in approximate inverse proportion. The dashed line in the
bottom panel of Fig. \ref{kmodeaverages} plots a simple $R^{-3/2}$ curve,
indicating that the average wavenumber follows a power-law distribution with
radius, such that $k_{\mathrm{av}} \sim R^{-3/2}$, which remains constant with
varying mass ratio.  This is easily understood by noting that since the sound
speed $c_{\mathrm{s}}$ is approximately constant by construction,
Eq.(\ref{kuns}) indicates that $k \sim \Sigma \sim R^{-3/2}$.

\subsection{Mach number of the spiral modes}
\label{spiralthermo}

Returning briefly to the dispersion relation given in
Eqs.(\ref{dispersionrelationD}) and (\ref{dispersionrelation}), we note that
this is only strictly valid for infinitesimally thin discs.  As our
simulations are fully three dimensional, we require a correction to the
self-gravity term to account for this, and this is shown in
Eq.(\ref{3Ddisprel}) below \citep{Bertin00},
\begin{equation}
  m^{2}(\Omega_{\mathrm{p}} - \Omega)^{2} = c_{\mathrm{s}}^{2}k^{2} - \frac{2
    \pi G \Sigma |k|}{1 + |k|H} + \Omega^{2},
  \label{3Ddisprel}
\end{equation}
where we have also used the fact that our discs are approximately Keplerian,
and thus $\kappa \approx \Omega$.  The reduction factor of $1/(1 + |k|H)$
arises from the vertical dilution of the gravitational potential due to the
finite thickness $H$ of the disc
\citep{Bertin00,BinneyTremaine2e,Vandervoort70a}.  Using this finite-thickness
dispersion relation and our averaged values for $k$ and $m$ it is possible to
calculate a Doppler-shifted angular speed $|\Omega_{\mathrm{p}} - \Omega|$,
noting that the sign of $\Omega_{\mathrm{p}}-\Omega$ cannot be determined from
Eq.(\ref{3Ddisprel}).  Since the average radial wave-number is always very
close to the most unstable one, and since the disc is almost exactly
marginally stable, the resultant average pattern speed turns out to be always
very close to co-rotation (as can be seen in Fig. \ref{xi}). We quantify the
deviation of the pattern speed from co-rotation later in section
\ref{locality}.

\begin{figure}
  \centering
  \includegraphics[width=20pc]{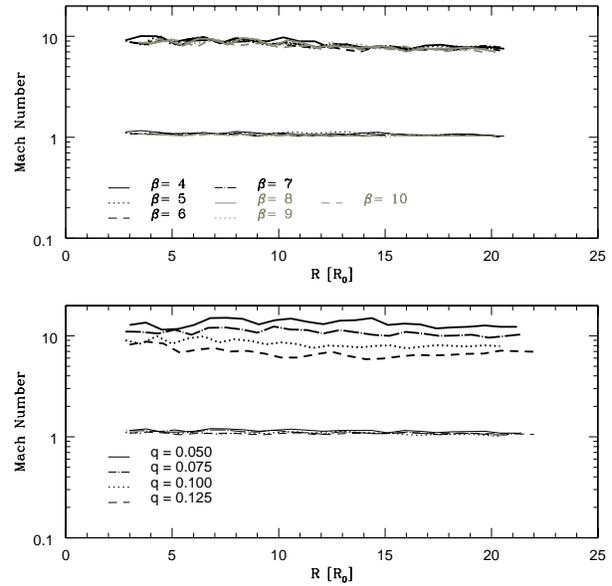}
  \caption{Wave phase Mach number $\mathcal{M}$ (thick lines) and the
    Doppler-shifted phase Mach number $\widetilde{\mathcal{M}}$ (thin
    lines) as a function of radius for various values of $\beta$ with $q =
    0.1$ (top) and as $q$ varies with $\beta = 5$ (bottom).} 
  \label{Machnumbers}
\end{figure}

We can further calculate the radial phase and Doppler-shifted phase Mach
numbers, and these are shown in Fig. \ref{Machnumbers}.  The upper panel shows
the wave phase Mach number $\mathcal{M}$ (thick lines) and the Doppler shifted
phase Mach number $\widetilde{\mathcal{M}}$ (thin lines) as functions of
radius for various values of $\beta$ with $q = 0.1$.  Similarly, the lower
panel of Fig. \ref{Machnumbers} shows the variation of these Mach numbers with
the mass ratio $q$ for $\beta = 5$.  We see immediately that both quantities
are independent of the cooling rate as measured by $\beta$ with very little
scatter.  Moreover, the Doppler-shifted phase Mach number is very close to
unity.  In a similar manner this quantity remains unchanged with variations in
the mass ratio, although the phase Mach number decreases with increasing $q$. 

\begin{figure}
  \centering
  \includegraphics[width=20pc]{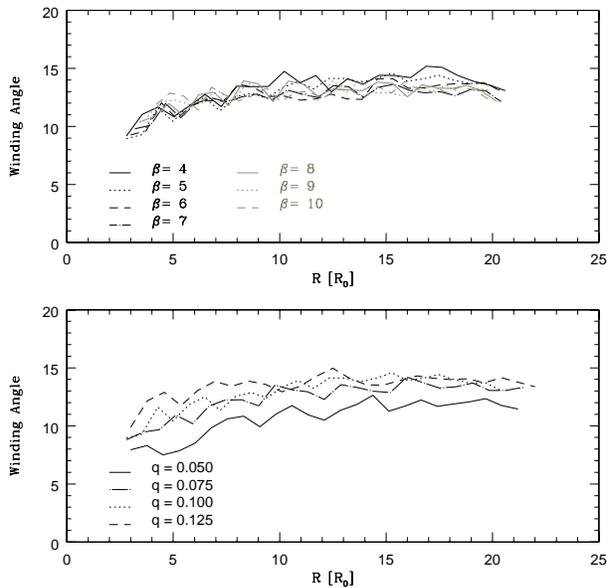}
  \caption{Wave winding angle $i$ as a function of radius plotted against
    varying cooling (top) and mass ratio (bottom).}
  \label{windingangle}
\end{figure}

The above results essentially imply that the wave structure is determined by
the requirement that the normal component of the flow into the shock is almost
exactly sonic -- a natural criterion for a quasi-steady system due to the
dissipative nature of shocks.  For waves with winding angle $i$, and radial
Doppler-shifted phase speed $\tilde{v}_{\mathrm{p}}$, a sonic normal component
of velocity into the shock implies $\tilde{v}_{\mathrm{p}} \cos i =
c_{\mathrm{s}}$, leading to  
\begin{equation}
  \widetilde{\mathcal{M}} = \frac{1}{\cos i}.
\end{equation}
Hence, in the limit of tightly wound waves where $\cos i \approx 1$, we should
expect that $\widetilde{\mathcal{M}} \approx 1$, as indeed we find in Figure
\ref{Machnumbers}.  For completeness, Fig. \ref{windingangle} shows the
winding angle $i$ as a function of radius for varying $\beta$ (top) and mass
ratio $q$, (bottom), using the definition $\tan i = m / kR$.  In all cases, $i
\lesssim 15^{\circ}$, so the waves are reasonably tightly wound
throughout. Again there is no significant variation with cooling, but the
structure becomes more open as the mass ratio increases, as expected from
Figs. \ref{rendered} and \ref{renderedq}.  

\begin{figure}
  \centering
  \includegraphics[width=20pc]{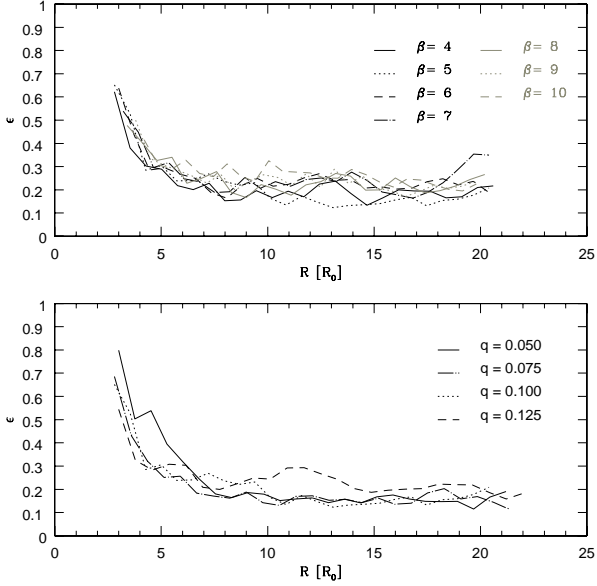}
  \caption{The heating factor $\epsilon$ as a function of radius for
    various values of $\beta$ with $q = 0.1$ (top) and as $q$ varies with
    $\beta=5$ (bottom).}
  \label{epsilon}
\end{figure}

We can also use Eq.(\ref{balance}) to estimate the amount of energy
dissipated by the weak spiral shocks per dynamical time, as characterised by
$\epsilon$, which is shown in Fig. \ref{epsilon}.  We have seen through the
constancy of the Doppler-shifted phase Mach number that the shock structure
that forms in the disc is indeed self-similar, and thus the heating factor
$\epsilon$ is also largely independent of the applied cooling, the mass ratio
and the radial position.  Note that the larger values for $\epsilon$
generated at low radii ($R \lesssim 5$) are probably due to the inaccuracies
in calculating $\delta \Sigma / \bar{\Sigma}$ in this region rather than a
breakdown in self-similarity. 

\subsection{On the locality of transport induced by self-gravity}
\label{locality}
\begin{figure}
  \centering
  \includegraphics[width=20pc]{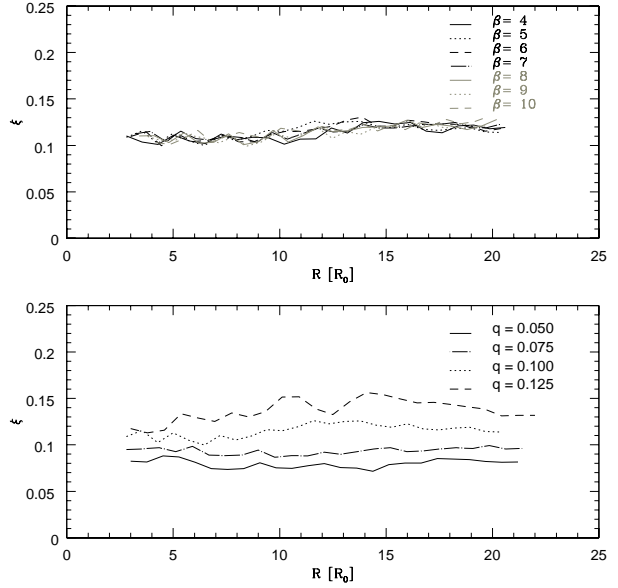}
  \caption{Non-local transport fraction $\xi$ as a function of
    radius for various values of $\beta$ with $q = 0.1$ (top) and as $q$ varies
    with $\beta = 5$ (bottom).}
  \label{xi}
\end{figure}

In the previous subsection we noted that the (spectrally averaged) pattern
speed of the waves $\Omega_{\mathrm{p}}$ is always very close to the angular
velocity of the flow $\Omega$, thereby indicating that the waves are close to
the co-rotation resonance as suggested earlier in the results of the radial
mode decomposition. We can estimate more quantitatively how close to co-rotation
the spiral waves lie by calculating the quantity $\xi$, as given in
Eq.(\ref{globaltransportfraction}).  This is shown in Fig. \ref{xi} as a
function of radius for all the values of $\beta$ and mass ratio simulated. 

For the $q=0.1$ case, we thus see that varying the cooling has no significant
effect on the transport properties of the disc, with $\xi \approx 0.1$
throughout the radial range, albeit with some scatter. This means that in this
configuration the disc is dominated by local transport processes, and is as
such reasonably well described by the viscous $\alpha$ prescription of
\citet{ShakSun73} -- global effects, although not negligible, are smaller than
local effects by an order of magnitude. 

By varying the mass ratio, we see that the strength of non-local effects
increases with $q$, rising to $\xi \approx 15\%$ for the case where $q =
0.125$. This confirms the results of \citet{LodatoR04}, who found similarly
that non-local effects (characterised by strong transient structures in the
disc) become increasingly important as the disc mass ratio rises, although for
the parameter range considered here the disc remains dominated by local
effects.  This non-local behaviour can be elucidated further by noting that
from the definitions of $\xi$ and $\tilde{v}_{\rm p}$
(Eqs.(\ref{globaltransportfraction}) and (\ref{Dopplerphasespeed})) we have    
\begin{equation}
  \xi \approx \widetilde{\mathcal{M}} \left( \frac{kH}{m} \right) 
\end{equation}
which, for $kH\approx 1$, reduces to
\begin{equation}
  \xi \approx \frac{\widetilde{\mathcal{M}}}{m} = \frac{1}{m \cos i}.
  \label{xi2}
\end{equation}
The non-locality of the transport is therefore directly linked to the openness
of the structure that is induced in the disc through self-gravity. Since, as
we have noted earlier, for larger disc masses the spiral structure tends to
become more open and more dominated by low $m$ modes, we then see that more
massive discs tend to become more subject to non-local effects.


\section{Discussion and Conclusions}
\label{discussion}

In this paper we have undertaken 3D global numerical simulations of gaseous,
non-magnetised discs, evolving under the influence of a massive central object
and their own self-gravity.  We have modelled the gas as an ideal gas with
$\gamma = 5/3$, together with a simple cooling prescription based on a local
cooling timescale.  We have used these simulations to investigate the
structure that forms once the discs have settled into a quasi-steady
marginally stable state as a function of both the imposed cooling and the disc
to central object mass ratio.

We have found that the amplitude of spiral arms induced in self-gravitating
discs, as characterised by the RMS surface density perturbations, can be
described straight-forwardly through the empirical relationship
\begin{equation}
  \left \langle \frac{\delta \Sigma}{\Sigma} \right \rangle \approx
  \frac{1.0}{\sqrt{\beta}},
\end{equation}
(where $\beta$ is the ratio between the local cooling and dynamical
timescales), with only a weak dependence on the disc to central object mass
ratio.  This is in fact closely linked to the result that the Doppler-shited
Mach number is very close to unity -- by considering the entropy change
$\Delta S$ across an adiabatic shock where the Mach number $M \approx 1$, it
can be shown that $\Delta S \sim (M^{2} - 1)^{2}$.  Thermal equilibrium in
these discs is established between cooling, at a rate inversely proportional
to $\beta$, and the irreversible conversion of mechanical energy into heat, at
a rate proportional to the entropy jump $\Delta S$ at the shock front -- we
therefore find that $\beta \sim (M^{2} - 1)^{-2}$.  Standard shock relations
show that the density perturbation $\delta \rho / \rho \sim (M^{2} - 1)$, and
hence simply from considering the properties of weak adiabatic shocks we can
arrive at the relationship $\delta \rho / \rho \sim \beta^{-1/2}$. 

Additionally we find that the heating factor $\epsilon$ -- that
fraction of the available wave energy that is liberated as heat back into the
disc gas -- remains essentially invariant at $\approx 20\%$ with both the
imposed cooling regime and the mass ratio of the disc to the central object.

As expected, our simulations show that the dominant radial wavenumber is
approximately equal to the reciprocal of the local scale height of the disc
throughout the radial range, $k \approx \pi G \Sigma / c_{\mathrm{s}}^{2}$.  We
therefore find that the radial spacing of the arms is dependent only on the
surface density and temperature profiles of the disc. Likewise although
further work is required to understand the relationship fully, the azimuthal
disc structure is dependent on the disc to central object mass ratio, with
more massive discs being characterised by more open structures than their
lower mass counterparts for a given central object mass.  

Our numerical results bear out the theoretical analysis of \citet{BalbusPap99}
and \citet{Gammie01}, who suggest that discs in the $Q \approx 1$ marginally
stable state may be modelled as predominantly local.  Simulations of
self-gravitating discs with radiative transfer by \citet{Boleyetal06} also
found that close to co-rotation, angular momentum transport was well modelled
by a local $\alpha$-prescription even when global modes were present. 
\citet{BalbusPap99} further predicted that non-local transport from an
``anomalous flux'' proportional to $\Omega - \Omega_{\mathrm{p}}$ would become
significant far from co-rotation, a result we have derived analytically using
the WKB approximation for tightly wound waves.  We have then used the WKB
dispersion relation along with empirically determined information on the
dominant wavenumbers to make an estimation of $|\Omega_{\mathrm{p}} -
\Omega|$. We find that, at least for low mass discs, this is a small fraction
of $\Omega$ (less than 15\% for discs with $q \leq 0.125$, regardless of the
efficacy of the cooling). Our results on the magnitude of the non-local
transport fraction $\xi = |\Omega_{\mathrm{p}} - \Omega| / \Omega$ can
furthermore be readily understood in terms of the empirical constancy of the
Doppler-shifted radial phase Mach number, $\widetilde{\mathcal{M}}$.  We
conclude that the importance of such non-local effects in gaseous
self-gravitating discs is set by the self-adjustment of the pattern speed to
ensure that the normal flow speed into the arms is sonic. We have then
demonstrated that this condition implies that $\xi \approx m^{-1} \sec i$,
where $i$ is the opening angle of the spiral structure.  Since the structure
within the disc becomes more open as the disc to central object mass ratio
increases, this implies that the importance of non-local transport also scales
with $q$. 

We note also that in collisionless sytems such as stellar discs, this
self-regulation process for the pattern speed breaks down as shocks cannot
form.  Hence it is possible to excite global modes in such discs, and thus
non-local transport of energy and angular momentum may be more significant
dynamically. The results that we present here are therefore restricted to the
case of predominantly collisional, gaseous discs.  Our results provide a
theoretical underpinning for the results of \citep{LodatoR04,LodatoR05} on how
the importance of global transport depends on the disc to central object mass
ratio in gaseous discs. In particular, we note that in cases (like those
described here) where the disc mass is a small fraction of the central object
mass (as could be the case for relatively evolved protostellar discs) the
effects of self-gravity are expected to be well described as a pseudo-viscous
process. 

One of the most important applications of our study is that we can relate the
amplitude of spiral modes in gaseous discs to the cooling regime.  With ALMA
coming online in the relatively near future, promising milli-arcsecond
resolution in the millimetre/sub-mm range, it is possible that such
observations of spiral structure in proto-planetary discs may become
technically feasible.


\section*{Acknowledgements}
\label{acknowledgements}

We acknowledge the use of SPLASH \citep{PriceSplash} throughout this paper for
the visualisation of surface densities.  We would also like to thank Jim
Pringle for helpful discussions and a careful reading of the manuscript.


\label{bib}
\bibliographystyle{mn2e} 
\bibliography{Cossins.bib}

\renewcommand{\theequation}{A-\arabic{equation}}
\setcounter{equation}{0}  
\section*{Appendix A: Resolution and Convergence Tests}
In this appendix we shall briefly outline the tests that were undertaken to
ensure the convergence of our results.

\begin{figure}
  \centering
  \includegraphics[width=20pc]{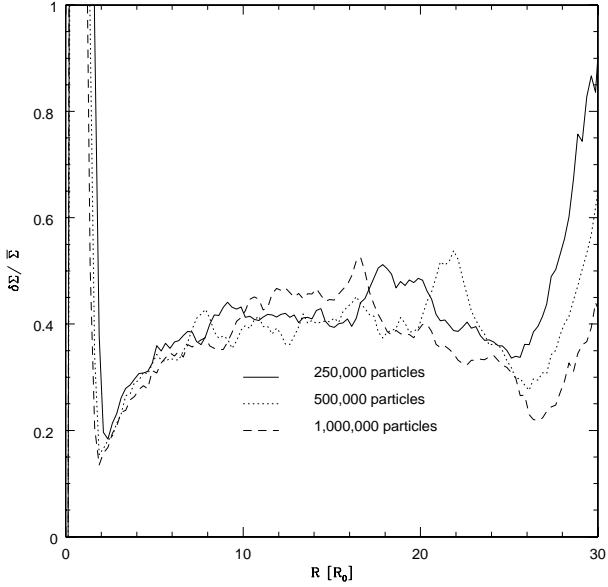}
  \caption{Variation of the radial RMS surface density perturbation amplitude
    as a function of simulation resolution.}
  \label{resolution_dsos}
\end{figure}

Three simulations were run, all with the cooling parameter $\beta = 6$ and
mass ratio $q = 0.1$, using discs of 250,000, 500,000 and 1,000,000 particles.
These were otherwise identical to the simulations that were used for this
paper, as described in full in section \ref{setup}.  The three values that are
of most significance to our results are the RMS surface density perturbation
amplitude $\delta \Sigma / \bar{\Sigma}$ and the average radial and azimuthal
wavenumbers $k_{\mathrm{av}}$ and $m_{\mathrm{av}}$ respectively.  Fig.
\ref{resolution_dsos} shows how $\delta \Sigma /\bar{\Sigma}$ varies with
resolution, and although there is considerable scatter it is clear that there
is no systematic variation with resolution. A similar result (with even less
scatter) is also obtained when one conducts a Fourier analysis of the
simulations -- there is no systematic variation with resolution.  We may
therefore conclude that our simulations are converged, and that the resolution
when using 500,000 particles is satisfactory for our purposes.

\section*{Appendix B: Fourier Decomposition Methods}
In this appendix we detail how the Fourier mode analysis was conducted, using
the SPH particle positions as the input values.  For simplicity, we begin by
discussing how the radial $k$ mode amplitudes were computed, as this had
practical implications on how the azimuthal $m$ mode analysis.

\subsection*{Radial mode analysis}
To calculate the radial Fourier mode amplitudes within a disc presents
certain problems, since even a cursory glance at Fig. \ref{rendered} reveals
that the radial wavenumber $k$ varies significantly with radius.  Also, unlike
the azimuthal wavenumbers, the disc is neither uniform nor periodic in radius,
and therefore the underlying Fourier distribution corresponding to the disc
surface density profile also has to be taken into account.  The following
method addresses both of these problems while keeping the signal to noise ratio
as high as possible. 

The disc to be analysed is divided into numerous overlapping annuli of width
$\Delta R$, which varies with the central radius of the annulus, and into a
number of sectors of fixed angular width $\Delta \phi$.  The $\Delta R$ values
are chosen such that each annulus is of sufficient radial extent to resolve
the greatest radial wavelength present at that radius, likewise each sector
must be narrow enough to ensure the wave crests are distinct and not smeared
out across a wide range in $R$.  In this manner, the smaller the winding angle
$\theta = \tan^{-1} |m/kR|$ of the waves the wider the sectors can be for a
given resolution.   

Since the radial wavenumber profile depends on the disc to central object mass
ratio, the radial extent $\Delta R$ of the annuli varied likewise, in order to
capture all the relevant modes.  The values used in our analyses are
summarised in Table \ref{kmodedata}.  Note that the widths of the annuli
increase linearly across the disc, from the initial to the final widths
quoted.      
\begin{table}  
  \begin{center}
    \begin{tabular}{ccccc}
      \hline 
      $M_{\mathrm{disc}}/M_{*}$ & Annuli & Initial width & Final width &
      Sectors \\
      \hline
      0.050 & 25 & 2 & 8  & 60 \\
      0.075 & 25 & 2 & 8  & 60 \\
      0.100 & 25 & 2 & 10 & 60 \\
      0.125 & 25 & 2 & 10 & 60 \\
      \hline
    \end{tabular}
    \caption{Details of the Fourier analyses for the various disc to central
      object mass ratios analysed.} 
    \label{kmodedata}
  \end{center}
\end{table}

To calculate the underlying Fourier distribution due to the unperturbed
surface density profile, the Fourier transform was taken over the whole of
each annulus.  This thereby smears out all the waves and takes the
\textit{average} distribution, and is evaluated according to the following
relation; 
\begin{equation}
  A_{\mathrm{k}} = \frac{1}{N_{\mathrm{ann}}} \left| \sum_{i
  = 1}^{N_{\mathrm{ann}}} e^{-i  k R_{i}} \right|,
  \label{kdisc}
\end{equation}
where $A_{\mathrm{k}}$ is the $k$ mode amplitude corresponding to
the underlying disc distribution, $N_{\mathrm{ann}}$ is the number of
particles per annulus, $k$ is the radial wavenumber and the $R_{i}$ are the
radii of the individual particles.  

The Fourier distribution of the waves overlaid on the disc are calculated by
taking an equivalent transformation over each sector within the annulus, such
that   
\begin{equation}
  A_{\mathrm{k,n}} = \frac{1}{N_{\mathrm{sect}}} \left|
  \sum_{i = 1} ^{N_{\mathrm{sect}}} e^{-i  k R_{i}} \right|,
  \label{kdiscwave}
\end{equation}
where $A_{\mathrm{k,n}}$ is the $k$ mode amplitude of the waves and disc
evaluated in the $n$th sector, and $N_{\mathrm{sect}}$ is the number of
particles in that sector.

Finally the Fourier distribution due solely to the waves in each sector is
given by the difference between equations (\ref{kdiscwave}) and (\ref{kdisc}). 
Since each sector should be statistically similar to the others we may then
average over all the sectors $N_{\mathrm{sectors}}$ (which in this case does
not smear the wave component out, but reduces computational noise), to give
the average radial Fourier mode amplitudes of the waves $\langle A_{k}
\rangle$;   
\begin{equation}
  \langle A_{k} \rangle = \frac{1}{N_{\mathrm{sectors}}} \sum_{n = 1}
  ^{N_{\mathrm{sectors}}} (A_{\mathrm{k,n}} - A_{\mathrm{k}}).
  \label{kwave}
\end{equation}

\begin{figure}
  \begin{center}
    \includegraphics[width=20pc]{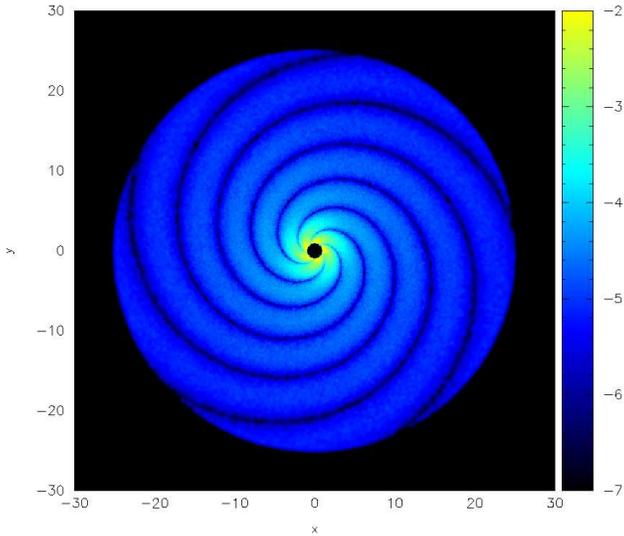}
    \caption{Test case for Fourier analysis showing the imposed structure.
      The colour scale shows logarithmic surface density.}
    \label{Sigma_test}
  \end{center}
\end{figure}

\subsection*{Azimuthal mode analysis}
For the azimuthal $m$ wavenumbers the analysis is more straightforward.  The
disc was initially divided into annuli of fixed width $\Delta R$ in such a
manner that each of these annuli is narrow enough to ensure the wave crests
occupy only a small range in $\phi$.  In contrast to the radial modes, these
annuli therefore need to become narrower with decreasing winding angle to
maintain resolution.  We found $\Delta R = 0.2$ (in code units) to be sufficient
for the purposes of this analysis. The azimuthal wavenumber amplitudes $A_{m}$
within each annulus are then computed via   
\begin{equation}
  A_{m} = \frac{1}{N_{\mathrm{ann}}} \left| \sum_{i = 1}^{N_{\mathrm{ann}}}
  e^{-im\phi_{i}} \right |, 
  \label{mmodes}
\end{equation}
where the $\phi_{i}$ are the azimuthal angles of the individual particles,
$N_{\mathrm{ann}}$ the number of particles in each annulus  and $m$ the radial
wavenumber of the wave, corresponding to the number of arms in the spiral. 

However, to ensure that we have the azimuthal $m$-mode amplitudes specified at
the same radii as the radial $k$-modes, then an average value is taken of the
$m$-mode amplitudes over all annuli where the central radius falls within that
annulus in which the $k$-modes are determined.

\begin{figure}
  \begin{center}
    \includegraphics[width=20pc]{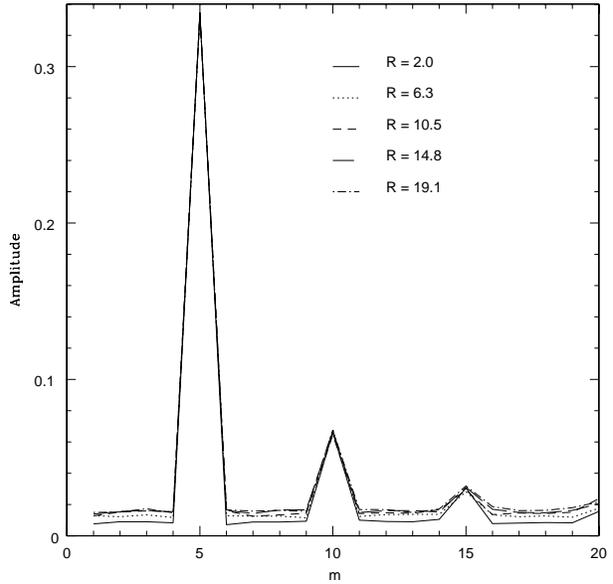}
    \caption{Results of the Fourier decomposition of test disc shown in
      Fig. \ref{Sigma_test} in terms of the azimuthal wavenumber, $m$.}
    \label{mtest}
  \end{center}
\end{figure}

\subsection*{Analysis Checks}
To ensure that the results of the Fourier analysis are accurate, we ran the
following test case.  A disc with an underlying surface density profile
$\Sigma \sim R^{-3/2}$ and an analytically superimposed structure was
created with five spiral arms, such that $m=5$ throughout the entire disc and
with the radial wavelength increasing linearly from $\lambda_{\mathrm{min}} =
2$ to $\lambda_{\mathrm{max}} = 7.6$.  This gives a total of five full
wavelengths across the face of the disc, which extends from $R = 1$ to $R =
25$.  The surface density of the disc, clearly indicating the imposed structure,
is shown in Fig. \ref{Sigma_test}.  The Fourier analysis was then conducted
using the annulus widths quoted in Table \ref{Ftest}, where as described above
the width of the annuli increased linearly from the minimum to the maximum
quoted value.   
\begin{table}  
  \begin{center}
    \begin{tabular}{cccc}
      \hline 
       Annuli & Initial width & Final width & Sectors \\
      \hline
      10 & 2.0 & 7.6  & 60 \\
      10 & 1.5 & 6.5  & 60 \\
      10 & 3.0 & 8.5  & 60 \\
      10 & 4.5 & 10.0 & 60 \\
      10 & 6.0 & 12.5 & 60 \\
      \hline
    \end{tabular}
    \caption{Details of the Fourier analyses for the test case}
    \label{Ftest}
  \end{center}
\end{table}

The results from the azimuthal Fourier decomposition are shown in
Fig. \ref{mtest}, and show that the azimuthal wavenumber is resolved extremely
well.  The fundamental frequency $m=5$ is clearly dominant, with no other
modes except higher harmonics present at any significant amplitude.  Note that
the results for the azimuthal modes show no sensitivity to the annuli used for
the analysis, and are quoted for the first case in Table \ref{Ftest} where the
annuli width correspond exactly to the radial wavelengths.

\begin{figure}
  \begin{center}
    \includegraphics[width=20pc]{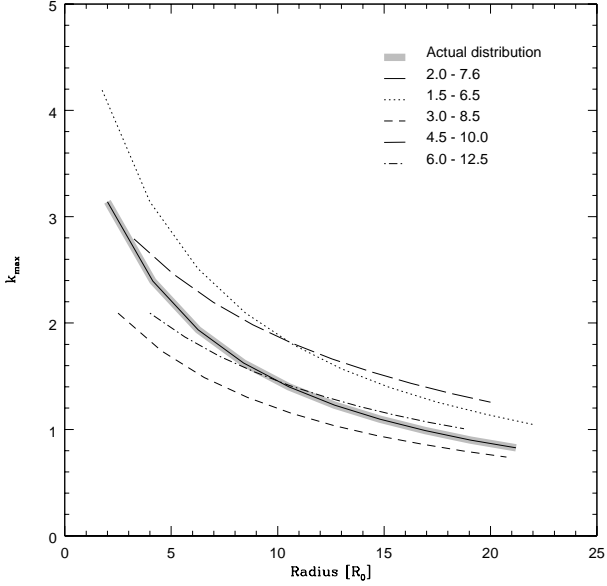}
    \caption{Results of the Fourier decomposition of the test disc shown in
      Fig. \ref{Sigma_test} showing the peak radial wavenumber
      $k_{\mathrm{max}}$ as a function of radius.   Various analyses are shown
      using the annuli given in Table \ref{Ftest}.}
    \label{ktest}
  \end{center}
\end{figure}

The results of the Fourier decomposition for the radial wavenumbers are shown
in Fig. \ref{ktest}, which shows the actual distribution of wavenumbers (as
calculated directly from the known distribution of wavelengths) and the
distributions derived from analyses using the annuli given in Table
\ref{Ftest}.   Note that the analysis using annuli that fit the wavelengths
exactly correspondingly reproduces the exact result.  Clearly there is scatter
within the results for the radial wavenumbers, which arises from the fact that
if the actual wavelength is not an integer divisor of the annulus over which
the analysis is being conducted, more than one wavenumber appears to be
excited. We note however that the scatter is never more than a factor of 1.5
above or below the true value, which we deem to be accurate enough for the
purposes of this analysis.  

\subsection*{Resolution Limits}
\begin{figure}
  \begin{center}
    \includegraphics[width=20pc]{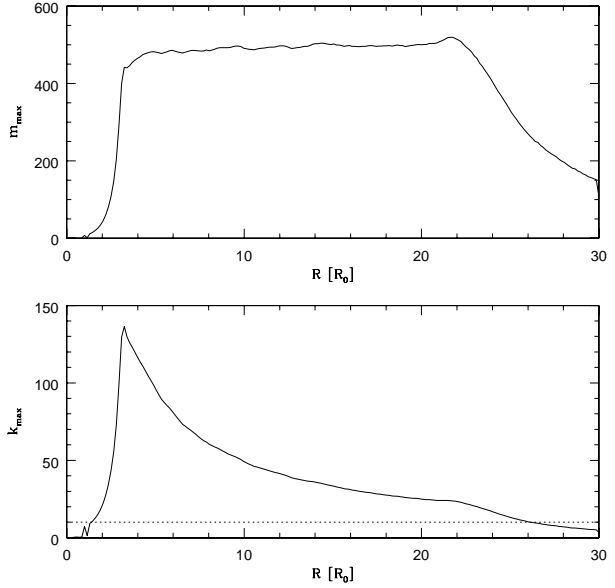}
    \caption{Resolution limits for the Fourier analysis in terms of the
      azimuthal wavenumbers (top) and radial wavenumbers (bottom).  The dashed
      line in the bottom plot indicates $k_{\mathrm{max}} = 10$.  The simulation
      parameters are $\beta = 10$, $q = 0.1$.}
    \label{resolutionlimits}
  \end{center}
\end{figure}

Throughout this paper, we have used discs of 500,000 particles when
undertaking the Fourier analysis.  From the fundamental SPH resolution limit
of the local smoothing length $h$ we can evaluate the maximum resolvable
azimuthal and radial wavenumbers $m_{\mathrm{max}}$ and $k_{\mathrm{max}}$ as
a function of radius throughout our discs, such that   
\begin{equation}
  m_{\mathrm{max}} = \frac{2 \pi R}{h} = 2 \pi \left( \frac{R}{H} \right)
  \left( \frac{H}{h} \right),
  \label{mlimit}
\end{equation}
\begin{equation}
  k_{\mathrm{max}} = \frac{2 \pi}{h} = \frac{2 \pi}{H} \left( \frac{H}{h}
  \right).
  \label{klimit}
\end{equation}
Since the approximate expected values for radial and azimuthal wavenumbers are
such that $kH \approx 1$ and $mH/R \approx 1$ respectively, Eqs.(\ref{mlimit})
and (\ref{klimit}) show that the accuracy of the Fourier analysis is closely
tied to the vertical resolution of the disc through $H/h$.  Using the average
smoothing length at each radius, these resolution limits are shown in
Fig. \ref{resolutionlimits}.  We have used data from the simulation where
$\beta = 10$, as this gives the most conservative limits of all our experiments.

\begin{figure}
  \centering
  \includegraphics[width=20pc]{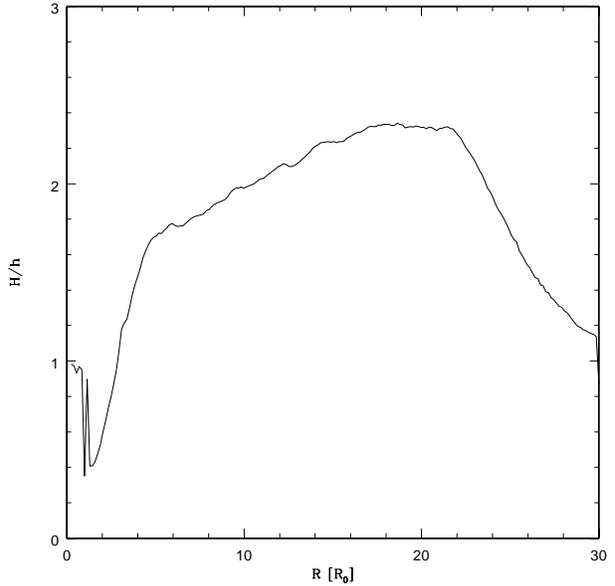}
  \caption{Ratio of the disc scale thickness $H$ to the average smoothing
    length $h$ as a function of radius.  Again the simulation parameters are
    $\beta = 10$, $q = 0.1$.} 
  \label{Hoverh}
\end{figure}

The vertical resolution of the disc as indicated by $H/h$ is shown in
Fig. \ref{Hoverh} for simulations using 500,000 particles.  We find that the
disc thickness is covered by approximately two smoothing lengths throughout, and
thus is adequately resolved.  For the Fourier analysis we therefore see that
the expected peak wavenumbers are resolved by a factor of approximately $4 \pi$
throughout the radial range.  For the radial wavenumbers, we are primarily
interested in $k < 10$, which is well resolved until at least $R = 25$, again
adequate for the analyses we have undertaken.  We conclude therefore that
throughout the radial ranges of interest, the Fourier analyses we have
presented are well resolved. 

\end{document}